\documentclass[useAMS,usenatbib]{mnras}
   
\usepackage{graphicx}
\usepackage{aasmacros}
\usepackage{color}
\usepackage{amsmath}
\usepackage{enumitem}

\def\bb{$\bullet$}

\def\hmpc{$h^{-1}$Mpc}
\def\hkpc{$h^{-1}$kpc}

\def\msol{M$_\odot$}

\def\rvir{R_{\rm vir}}

\def\s8{\sigma_8}

\def\lcdm{$\Lambda$CDM}

\def\x2{$\chi^2$}

\def\NNm1{\langle N(N-1) \rangle}

\def\m_star{M_\ast}

\def\lcdm{$\Lambda$CDM}

\def\s8{\sigma_8}

\def\hmpc{$h^{-1}\,$Mpc}
\def\hkpc{h^{-1}{\rm kpc}}

\def\x2{$\chi^2$}

\def\NNm1{\langle N(N-1) \rangle}

\def\p0{P_0(r)}

\def\mgal{{M_{\ast}}}

\def\mhalo{M_{h}}
\def\lesssim{\la}

\def\zhalf{z_{1/2}}
\def\psat{P_{\rm sat}}
\def\psatiso{P_{\rm cen}}
\def\pcen{P_{\rm cen}}

\def\msol{{\rm M}_\odot}

\def\lsat{L_{\rm sat}}
\def\mpeak{M_{\rm peak}}
\def\rvir{R_{\rm vir}}
\def\zhalf{z_{1/2}}
\def\lsatx{L_{\rm sat}^{50}}
\def\lsatxx{L_{\rm sat}^{100}}
\def\lsatr{L_{\rm sat}^R}
\def\lbg{L_{\rm BG}}
\def\ltot{L_{\rm tot}}
\def\phisat{\Phi_{\rm sat}}
\def\hkpc{h^{-1}{\rm kpc}}
\def\vpeak{V_{\rm peak}}
\def\hmpc{h^{-1}{\rm Mpc}}
\def\lcdm{\Lambda{\rm CDM}}

\title[Total satellite luminosity]{Probing the galaxy-halo connection
  with total satellite luminosity}
\author[Tinker et.~al.]{\parbox{\textwidth}{Jeremy
    L. Tinker$^1$, Junzhi Cao$^1$, Mehmet Alpaslan$^1$, Joseph DeRose$^{2,3,4,5,6}$,\\
    Yao-Yuan Mao$^{7,8,9}$, Risa H. Wechsler$^{2,3}$}\\
  \footnotesize
  \\
  $^1$Center for Cosmology and Particle Physics, Department of
  Physics, New York University, New York, NY, 10003, USA\\
$^2$Kavli Institute for Particle Astrophysics and Cosmology and
Department of Physics, Stanford University, Stanford, CA 94305, USA\\
$^3$Department of Particle Physics and Astrophysics, SLAC National
Accelerator Laboratory, Stanford, CA 94305, USA\\
$^4$Santa Cruz Institute for Particle Physics, Santa Cruz, CA 95064, USA\\
$^5$Berkeley Center for Cosmological Physics, Department of Physics, University of California, Berkeley CA 94720\\
$^6$Lawrence Berkeley National Laboratory, 1 Cyclotron Road, Berkeley, CA 93720\\
$^{7}$Department of Physics and Astronomy, Rutgers, The State University of New Jersey, Piscataway, NJ 08854, USA\\
$^{8}$Department of Physics and Astronomy, University of Pittsburgh, Pittsburgh, PA 15260, USA\\
$^{9}$Pittsburgh Particle Physics, Astrophysics, and Cosmology Center (PITT PACC), University of Pittsburgh, Pittsburgh, PA 15260, USA\\
}

\hypersetup{draft}
\begin{document}
\label{firstpage}
\pagerange{\pageref{firstpage}--\pageref{lastpage}}
\maketitle

\begin{abstract}

  We demonstrate how the total luminosity in satellite galaxies is a
  powerful probe of dark matter halos around central galaxies. The
  method cross-correlates central galaxies in spectroscopic galaxy
  samples with fainter galaxies detected in photometric surveys. After
  background subtraction, the excess galaxies around the spectroscopic
  central galaxies represent faint satellite galaxies within the dark
  matter halo. Using abundance matching models, we show that the the
  total galaxy luminosity, $\lsat$, scales linearly with host halo
  mass, making $\lsat$ an excellent proxy for $\mhalo$. $\lsat$ is
  also sensitive to the formation time of the halo, as younger halos
  have more substructure at fixed $\mhalo$. We demonstrate that probes
  of galaxy large-scale environment can break this degeneracy in
  $\lsat$. Although this is an indirect measure of the halo, it can
  yield a high signal-to-noise measurement for galaxies expected to
  occupy halos at $<10^{12}$ $\msol$, where other methods suffer from
  larger errors.  In this paper we focus on observational and
  theoretical systematics in the $\lsat$ method.  We test the
  robustness of our method of finding central galaxies in
  spectroscopic data and our methods of estimating the number of
  background galaxies.  We implement this method on central galaxies
  and galaxy groups identified in SDSS spectroscopic data, with
  satellites identified in faint imaging from the DESI Legacy Imaging
  Surveys. We find excellent agreement between our abundance matching
  predictions and the observational measurements, confirming that this
  approach can be used as a proxy for halo mass. Finally, we compare
  our $\lsat$ measurements to weak lensing estimates of $\mhalo$ for
  red and blue subsamples of central galaxies. In the stellar mass
  range where the measurements overlap, we find consistent results,
  where red galaxies live in larger halos than blue galaxies. However,
  the $\lsat$ approach allows us to probe significantly lower mass
  galaxies. At these lower stellar masses, the $\lsat$ values---and,
  by extension, the halo masses---are equivalent. This example shows
  the potential of $\lsat$ as complementary to weak lensing as a probe
  of dark matter halos.

\end{abstract}


\section{Introduction}

The connection between galaxies and dark matter halos is critical both
for our understanding of galaxy formation and for constraining
cosmology (see \citealt{wechsler_tinker:18} for a recent
review). There are myriad approaches to observationally constraining
the galaxy-halo connection. These approaches generally separate into
two categories: indirectly inferring the connection statistically, and
directly probing the dark matter halo masses through their
gravitational potential. In this paper, we develop a method that is
complementary to, but distinct from, these direct approaches. This
method, which uses the total luminosity in satellite galaxies,
$\lsat$, is a direct observable, but it is an indirect probe of halo
mass. However, it is yields a higher signal-to-noise measurement than
other methods and is robust to significantly lower mass dark matter
halos. Using the luminosity in satellite galaxies only, as opposed to
including the light from the central galaxy, makes the approach far
more sensitive to the properties of halos around galaxies at the Milky
Way mass scale and below.

Indirect, statistical approaches to quantifying the galaxy-halo
connection usually focus on galaxy clustering and number densities
(e.g., \citealt{zehavi_etal:05, zehavi_etal:11, tinker_etal:07_pvd,
  zheng_etal:07, rodriguez_puebla_etal:15}) or from galaxy group
catalogs (\citealt{yang_etal:08, reddick_etal:13,
  sinha_etal:18}). These analyses parameterize the relationship
between galaxies and halos with a halo occupation distribution (HOD)
and then constrain the free parameters from the data.  \footnote{One
  can also use hybrid approaches that incorporate satellite kinematic
  data or weak lensing measurements with clustering and abundances to
  constrain the free parameters of the halo occupation models
  (\citealt{more_etal:11, leauthaud_etal:12_shmr, tinker_etal:13,
    zu_mandelbaum:15, lange_etal:18}). But these hybrid approaches are
  still distinct from direct approaches that only use observables
  sensitive to the dark matter potential, without any constraints
  based on halo occupation models and the number density of galaxies.}
However, these indirect methods are most effective for halo mass
scales above $\mhalo\approx 10^{12}$ $\msol$, or for galaxies above
$\mgal\approx 10^{10.5}$ $\msol$. In these clustering-based
approaches, halo clustering becomes independent of mass at
$\mhalo<10^{12}$ $\msol$, thus the observed clustering amplitude
carries little information about the host halo masses of central
galaxies below these mass scales. This lack of constraining power is
evident when comparing different analyses of low-redshift SDSS galaxy
samples to determine the galaxy-halo connection for red and blue
galaxy subsamples. \cite{wechsler_tinker:18} compiled recent results
of the relative stellar masses of red and blue central galaxies at
fixed halo mass (cf. their Figure 8, which presents
$\mgal_{\rm red}/\mgal_{\rm blue}$ as a function of $\mhalo$). Below
$10^{12}$ $\msol$, there is nearly an order of magnitude difference in
$\mgal_{\rm red}/\mgal_{\rm blue}$, even though most of the analyses
utilize the same SDSS dataset. Due to the lack of constraining power
in the data, the results reflect the assumptions made in the modeling
rather than physical reality.

Direct probes of the gravitational potential of dark halos are through
gravitational lensing and satellite kinematics (e.g., \citealt{
  conroy_etal:05, norberg_etal:08, hudson_etal:15,
  mandelbaum_etal:16}). For direct probes, when applied to the large
spectroscopic surveys with the SDSS, the signal-to-noise of direct
probes degrades rapidly at $\mhalo < 10^{12}$ $\msol$. Deeper and
higher resolution imaging makes measurements of lensing masses tenable
at lower masses, but at the cost of sample size.

Thus, there is a need for a complementary method of probing dark
matter halos around lower mass galaxies; one that is a direct
observable, uninfluenced by model assumptions of halo occupation
methods but has the statistical precision to probe the halos around
low-mass galaxies. The halo mass scale of $10^{12}$ $\msol$ is
auspicious for studies of galaxy formation. The
transition from gas accretion being predominantly cold to exclusively
hot occurs at or below this scale (\citealt{dekel_birnboim:06,
  keres_etal:05, keres_etal:09}). From abundance matching analyses,
this is also the scale at which most star formation occurs throughout
the history of the universe (\citealt{behroozi_etal:13_letter}). As a
result, all approaches to constrain the galaxy halo connection find
that that this mass scale is the {\it pivot point} of the
stella-to-halo mass relation, where $\mgal/\mhalo$ is maximal (see
\citealt{wechsler_tinker:18} and references therein). Thus, being able
to directly probe the relationship between galaxies and halos at
scales of $\mhalo \sim 10^{11}-10^{12}$ $\msol$ would open a door to
our understanding of galaxy formation that has to this point remained
closed.

The $\lsat$ method probes dark matter halos by measuring the total
amount of light in satellite galaxies within a dark halo. All galactic
dark matter halos, regardless of their mass, contain significant
amounts of substructure within them. These substructures, which we
refer to as subhalos, also can contain galaxies. Simulations show that
the number of galaxy-occupied subhalos should scale roughly linearly
with host halo mass (e.g., \citealt{kravtsov_etal:04,
  reddick_etal:13}). For a spectroscopic
survey like SDSS, the majority of satellites in Milky Way-type halos
are significantly below the magnitude limit to be selected for
spectroscopy. Recently, the SAGA survey (\citealt{saga}) performed a
detailed study of 8 Milky-Way like galaxies, searching for satellite
galaxies at significantly fainter magnitudes. Each galaxy contained
faint satellites, with the number ranging from 2 to 10 objects. We use
a complementary approach here: rather than assign specific satellites
to specific systems, we stack sets of spectroscopic galaxies to
measure the mean number of faint satellites within the
stacked sample, detected in deeper imaging data.

More than constraining the mean relationship between $\mgal$ and
$\mhalo$, the $\lsat$ method we present here can uncover secondary
trends in the galaxy-halo connection. Because $\lsat$ is not a direct
observable of the the gravitational potential, an independent
calibration would be required to convert $\lsat$ to $\mhalo$. However,
comparing the relative values of $\lsat$ of galaxies is a robust
observable, and presents a unique test for galaxy formation
models. For example, as presented in \cite{wechsler_tinker:18}, it is
an open question whether the scatter in $\mhalo$ at fixed $\mgal$
correlates with any other galaxy or halo properties, such as galaxy
size, stellar velocity dispersion, or morphology. In a companion paper
(\citealt{alpaslan_tinker:19}) we present first results quantifying these
correlations using the $\lsat$ method.

An important caveat with this method is that, when measuring relative
values of $\lsat$, one is not necessary measuring relative values of
$\mhalo$. It is well know that the amount of substructure in a halo is
correlated with the formation history of that halo, an effect that is
part of `halo assembly bias' (e.g., \citealt{wechsler_etal:06,
  gao_white:07, mao_etal:15}). This makes $\lsat$ a {\it more}
insightful probe of the galaxy-halo connection than simply probing
dark matter mass itself. Ever since the first discoveries of halo
assembly bias, the persistent and controversial question has been
whether this effect propagates into the galaxy formation process (see
the discussion in \citealt{wechsler_tinker:18}). If secondary
properties of central galaxies correlate with halo formation history,
these properties will correlate with large-scale environment in a
distinct manner. The $\lsat$ observable, combined with these spatial
tests (e.g., \citealt{abbas_sheth:06, tinker_etal:08_voids,
  peng_etal:10, tinker_etal:17_p1, tinker_etal:18_p3,
  zu_mandelbaum:18, walsh_tinker:19, wang_etal:19}), offers to ability
not only to tell if central galaxies of the same mass, but different
secondary properties, live in different halos, but can discriminate
between differences in halo mass or in halo formation history.

In this paper, we present the following:

\begin{itemize}[leftmargin=0.7cm]
\item Theoretical predictions for $\lsat$.
\item Tests of our observational methods.
\item First results of $\lsat$ vs $\mhalo$ and $\lsat$ vs $\mgal$.
\end{itemize}

\noindent The method is based on stacking objects of a common set of
properties. The objects the stacks are centered on---spectroscopic
galaxies with a likelihood of being central galaxies within their host
halos--- we will refer to as `primary' objects. Once the stack is
collated, we measure the luminosity function of faint objects around
the primaries, assuming all galaxies are at the redshift of the
primary. We subtract off the estimated background contribution to this
luminosity function from objects projected along the line of
sight. Whatever remains after background subtraction are satellite
galaxies associated with the halos of the primary objects. This
technique has been used to quantify satellite populations in massive
objects, such as galaxy clusters (e.g., \citealt{hansen_etal:09}) and
luminous red galaxies (\citealt{tal_etal:12}). With these types of
objects, there is minimal contamination by misclassification of
satellite galaxies to be in the primary galaxy sample. Here we extend
this method to much fainter primary objects, using new methods to
minimize the contamination of satellite galaxies.

Our theoretical predictions are constructed using the abundance
matching framework (see, e.g., \citealt{wechsler_tinker:18}), coupled
with high-resolutions N-body simulations that resolve substructure
even within low-mass halos. The tests of the observational methods
focus on two aspects: defining a sample of central galaxies within the
spectroscopic galaxy sample, and estimating the density of background
imaging galaxies. For measurements on survey data, primary objects are
selected from the Main Galaxy Sample of the SDSS
(\citealt{strauss_etal:02}).  Secondary objects come from the DESI
Legacy Imaging Surveys (\citealt{legacy_surveys}), which has an
$r$-band depth of $r\sim 24$, more than 6 magnitudes fainter than the
SDSS spectroscopic sample. 

The outline of this paper is as follows: In \S 2, we detail the
observational data, both imaging and spectroscopic, that will be
utilized to make $\lsat$ measurements. In \S 3, we will construct
theoretical models for $\lsat$ from the abundance matching framework,
focusing not just on how $\lsat$ scales with $\mhalo$ but also how
other halo properties correlate with satellite luminosity. In \S 4
will evaluate the efficacy of our methods for finding central galaxies
in spectroscopic samples, but volume- and flux-limited. In \S 5, we
present details of applying the method to observational data,
presenting tests of the method on mock galaxy distributions. In \S 6,
we present our first results on the $\lsat$ of dark matter halos
around SDSS central galaxies. In \S 7 we summarize and discuss the
results.


\begin{table}
\caption{{\bf Volume-Limited SDSS Samples}. The first column sets the
  magnitude threshold at the maximum redshift, while the stellar mass
  listed is the limit of 98\% completeness. Galaxies must be brighter
  and more massive than the limits listed to be in the sample. Col. 4
  lists the minimum halo mass returned by the group finder.}
\label{t.samples}
\begin{tabular}{ccccc}
\hline
$M_r-5\log h$ & $z_{\rm max}$ & $\log \mgal/[\msol]$ &  Min. $\mhalo$ [$\msol$] &  $N_{\rm gal}$\\
\hline
-17.48 & 0.033 &  9.7 &$2.4\times 10^{11}$ & 16975 \\
-18.32 & 0.047 & 10.1 & $3.3\times 10^{11}$ & 30144 \\
-19.04 & 0.065 & 10.5 & $8.1\times 10^{11}$ & 41659 \\
\end{tabular}
\end{table}


\begin{table}
  \caption{N-body Simulations}
  \label{t.nbody}
\begin{tabular}{cccccccc}
\hline

Name & $\Omega_m$ & $\Omega_b$ & $\sigma_8$ & $H_0$ & $L_{\rm box} $ & $N_{\rm p}$ & $m_p$ \\
 \hline
  C250 & 0.295 & 0.047 & 0.834& 68.8 & 250& 2560$^3$& 7.63$\times 10^7$ \\ 
  C125 & 0.286 & 0.047 & 0.82& 70.0& 125& 2048$^3$& 1.8$\times 10^7$ \\ 
\end{tabular}
\end{table}

\section{Data}
\label{s.data}

Our analysis combines spectroscopic data from SDSS and imaging data
from the DESI Legacy Imaging Surveys. First, we will describe how we
construct samples of central galaxies from SDSS data. Central galaxies
come from both volume-limited catalogs as well as the full
flux-limited Sloan catalog. Second, we describe the imaging data,
including the cuts employed to create the sample of secondary
galaxies.

\subsection{SDSS Central Galaxies}

We use the spectroscopic data from the Sloan Digital Sky Survey Main
Galaxy Sample to find central galaxies
(\citealt{strauss_etal:02}). Specifically we use the DR7.2 release of
the NYU-VAGC catalogs (\citealt{blanton_etal:05_vagc}).

We use two complementary methods to construct samples of central
galaxies: volume-limited group catalogs, and a flux-limited catalog of
central galaxies. For the former, we use a combination of three
different volume-limited samples, listed in Table 1. This table also
shows the minimum halo mass in the sample, as determined by our galaxy
group catalog. The group-finding algorithm we use is detailed in
\cite{tinker_etal:11}, based on the halo-based group finder of
\cite{yang_etal:05}, and further vetted in
\cite{campbell_etal:15}. The standard implementation of the group
finder yields central galaxy samples with a purity and completeness of
$\sim 85-90\%$ (\citealt{tinker_etal:11}), but we will discuss
methods of using the group finder information to construct
higher-purity samples with limited loss of completeness.

We also apply a central-galaxy finding algorithm to the full
flux-limited SDSS main galaxy sample. The algorithm is described in
detail in Appendix \ref{s.app_cenfinder}. In short, the approach
relies of pre-tabulated stellar-to-halo mass relations (SHMRs) to
assign halos to observed galaxies, then uses the same probabilistic
approach as that used in the group catalog to determine whether a
given galaxy is a satellite within a larger halo. This method,
although approximate, yields central galaxy samples consistent with
that of the volume-limited group catalog. But the use of pre-tabulated
SHMRs does not require that the sample be
volume-limited. Additionally, it does not require the sample to be
statistically representative, which is a requirement of the group
catalog. This allows the central finder to yield robust results at
high redshifts, where only the brightest galaxies exist, and at low
stellar masses, which are only found in small numbers at the lowest
redshifts.

In this paper we will use different projected apertures within which
to measure $\lsat$. We define $\lsatr$ as the satellite luminosity
within the projected virial radius of the halo, $\rvir$. We define
$\lsatx$ and $\lsatxx$ as satellite luminosity within fixed comoving
projected apertures of 50 $\hkpc$ and 100 $\hkpc$ respectively.  To
measure $\lsatr$ from SDSS data, we use the volume-limited group
catalogs because the halo mass estimates are reasonably accurate. To
measure $\lsatx$ as a function of $\mgal$, we use the flux-limited
central catalog because it has better statistics at both low and high
stellar masses.

\subsection{Legacy Survey Imaging Data}

The DESI Legacy Imaging Surveys (DLIS) is a combination of three different imaging
surveys. At Declinations below +32$^\circ$, in both the NGC and SGC,
data come from the DECam instrument on the 4-meter Blanco telescope
(\citealt{decam}). This includes $g$, $r$, and $z$-band data,
referenced as the DECam Legacy Survey (DECaLS). In the SGC above
+32$^\circ$, data come from the Beijing-Arizona Sky Survey (BASS) on
the Bok telescope (\citealt{bass_mzls}). This includes $g$ and
$r$-band imaging. The $z$-band imaging over the same area of the sky
comes from the Mayall $z$-band Legacy Survey (MzLS;
\citealt{bass_mzls}). Once completed, the total area covered in the DLIS
will be 14,000 deg$^2$, with DECaLS comprising 9,000 deg$^2$ and the
combined northern facilities supplying the additional 5,000 deg$^2$ of
coverage to complete the footprint. The DLIS footprint is largely
coincident with the footprint of the final SDSS imaging footprint.

For this paper we use Data Releases 6 and 7. DR6 comprises the NGC
surveys, BASS and MzLS, while DR7 is the latest DECaLS release. DR6
covers 3,823 deg$^2$ with at least one pass in each imaging
band, and 1,441 deg$^2$ of three-pass coverage in all bands, which is
the full depth of the survey. DR7 covers nearly 9,300 deg$^2$ with at
least one pass in each imaging band, and 4,355 deg$^2$ of three-pass
coverage in all bands. Although our fiducial results use only $r$-band
data to measure $\lsat$, we require at least one pass in all three
bands for data to be part of our analysis. Single-pass depth of the
survey is $\sim 23.3$ in $r$-band, with a full 3-pass depth of $\sim
23.9$, although the exact depth of the survey fluctuates across the
footprint at fixed pass number. We will detail to quality cuts imposed
on the data in \S \ref{s.application}.

Before doing any measurements, we first perform quality cuts on the
data. Starting from the DLIS sweep files, we perform the following cuts
on the data:

\begin{description}
\item \bb\hspace{0.1cm} Remove all objects with ${\tt type} == `{\rm PSF}'$
\item \bb\hspace{0.1cm} Remove all objects with ${\tt nobjs== 0}$ in the $g$, $r$, or
  $z$ bands.
\item \bb\hspace{0.1cm} Remove all objects with ${\tt flux\_ivar<=0}$ in the $g$, $r$, or
    $z$ bands.
\item \bb\hspace{0.1cm} Remove all objects with ${\tt fracmasked>0.6}$ in the $g$, $r$, or
    $z$ bands.
\end{description}

\noindent The {\tt fracmasked} keyword specifies the total fraction of
pixels in an object that are masked out. Masking can be due to bright
stars, saturated pixels, and a number of other minor occurrences that
are detailed in the DLIS documentation. To account for the minor
differences in the geometry of the DLIS survey and the footprint of
the SDSS MGS, we only include SDSS central galaxies that lie far
enough within the DLIS footprint such that an annulus with $R=3\rvir$
of the estimated dark matter halo is completely within the survey area
covered by at least one observation in all of $g$, $z$, and $r$ bands.

\begin{figure*}
  \includegraphics[width=6in]{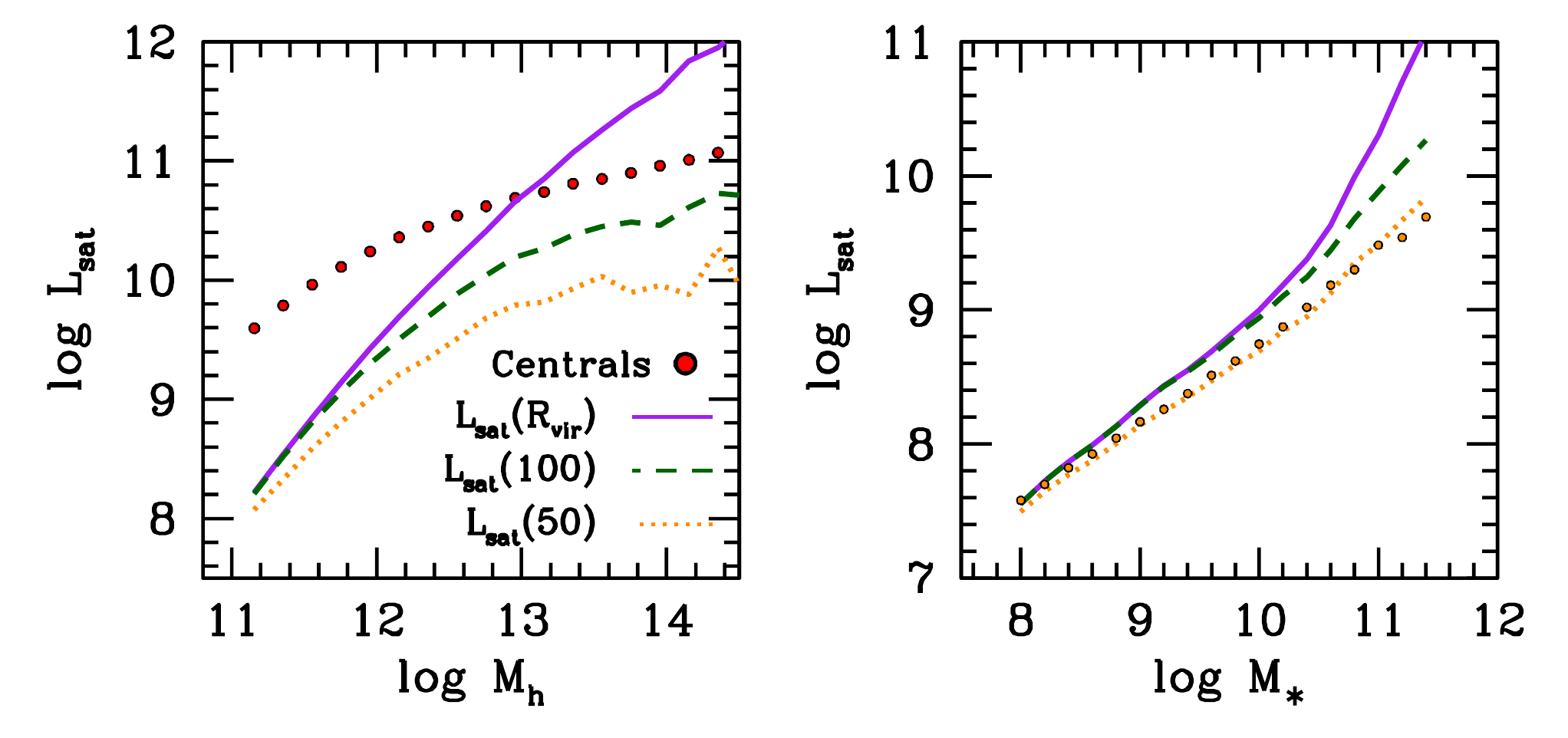}
  \vspace{-0.4cm}
  \caption{ \label{lsat_theory} {\it Left Panel:} Dependence of
    $\lsat$ on $\mhalo$ predicted by abundance matching applied to
    halos in the C125 simulation. $\lsat(\rvir)$ means all satellites
    within the virial radius, while 100 and 50 refer to $\lsat$ within
    projected apertures, in $\hkpc$. For comparison, the luminosity of
    central galaxies is shown with the filled circles. At
    $\mhalo\lesssim 10^{12}$, the luminosity of the central galaxy
    dominates the total luminosity within the halo. This is why using
    $\lsat$ is a far more sensitive diagnostic of the dark halo than
    using $\lsat+L_{\rm cen}$. {\it Right Panel:} Same as left panel,
    but now binning the object by $\mgal$ rather than $\mhalo$.  The
    filled circles show the results of a model in which the scatter of
    $\mgal$ at fixed $\mhalo$ is perfectly correlated with $\zhalf$,
    the formation time of the halo. Although $\lsat$ correlates with
    $\zhalf$, this does not impact results when binning in $\mgal$. }
\end{figure*}

\section{Theoretical Predictions for $\lsat$}
\label{s.theory}

Before testing the methodology of measuring $\lsat$ on mocks and data,
we first need a baseline expectation for $\lsat$ as a quantity that
correlates with dark matter halos. In this section, we present
our framework for constructing these theoretical models, and test not
just how $\lsat$ scales with $\mhalo$, but also how $\lsat$ scales with
$\mgal$ and how it correlates with secondary halo properties other
than mass.

\subsection{Numerical Simulations and Methods}

To make theoretical predictions for $\lsat$ in the context of $\lcdm$
structure formation, we combine high-resolution N-body simulations
with abundance matching models. Table \ref{t.nbody}
shows the properties of the two simulations usd to make predictions
here. Both use the ROCKSTAR code (\citealt{rockstar}) to identify
halos and ConsistentTrees (\citealt{consistent_trees}) to track the
merger histories of individual galaxies.

The smaller-volume simulation, C125, has $\sim 4$ times better mass
resolution than the larger-volume C250. We require that each
simulation have enough resolution to track the subhalos that would
contain satellites down to an absolute magnitude of $M_r-5\log
h=-14$, which is the current limits of our observational
results. Comparison between the two simulations at fixed host halo
mass shows a small offset in the total $\lsat$ values at
$\mhalo\lesssim 10^{12}$ $\msol$ of roughly 0.1-0.2
dex. We will use C125 to show predictions of the mean trends of
$\lsat$, but use C250 to predict clustering results because the larger
volume is necessary to the reduce noise in clustering results. 

To connect galaxy luminosity and stellar mass to dark matter halos, we
use abundance matching (see, e.g., \citealt{wechsler_tinker:18}). To
assign $M_r$ to halos and subhalos, we use the \cite{blanton_etal:05}
luminosity function, which specifically corrects for incompleteness in
SDSS observations of low-luminosity, low surface brightness
galaxies. We connect $M_r$ to $\mpeak$, the peak halo mass throughout
a (sub)halo's history assuming a scatter of 0.2 dex in luminosity at
fixed $\mpeak$\footnote{We note that there is marginal ($<0.1$ dex)
  difference in the results when using $\vpeak$---the largest value of
  the halo's maximum circular velocity during its evolution---as the
  halo parameter to abundance match to}. To assign $\mgal$ to halos
and subhalos, we use the stellar mass function presented in
\cite{cao_etal:19}, which utilized the PCA galaxy stellar masses of
\cite{chen_etal:12}. The stellar mass function for SDSS DR7 is also
presented in Cao et al. Although we require that our abundance
matching in luminosity is robust down to very low luminosities, for
stellar mass we only require that the abundance matching is accurate
down to $\mgal\sim 10^9$ $\msol$, which is the lower limit of our
sample of central galaxies in SDSS.

\subsection{Scaling of $\lsat$ with $\mhalo$ and $\mgal$}

Figure \ref{lsat_theory} shows the predictions for $\lsat$ from the
C125 simulation, binned both as a function of $\mhalo$ and
$\mgal$. Results are for all satellites brighter than
$M_r-5\log h<-14$. In both panels, the results are shown for our three
different apertures.

The left-hand side of Figure \ref{lsat_theory} shows the results as a
function of $\mhalo$. For $\lsat(\rvir)$, the total satellite
luminosity scales roughly as a power law, with a scaling slightly
steeper than linear. This is expected, as subhalo populations are
mostly self-similar when you scale up the host halo mass
(\citealt{gao_etal:04}), and the number of subhalos at fixed $\mpeak$
scales linearly with host halo mass
(\citealt{kravtsov_etal:04}). Between $\mhalo=10^{11}$ to $10^{12}$
$\msol$, the luminosity of the central galaxy increases by a factor of
$\sim 4$, while $\lsat$ increases by a factor of $\sim 20$, making is
a much more sensitive diagnostic of halo mass than either the
luminosity of the central galaxy or the combined luminosity of both
central and satellite galaxies together.

For the 50 and 100 $\hkpc$ apertures---$\lsatx$ and $\lsatxx$,
respectively---the results at $\mhalo<10^{12}$ $\msol$ are mostly
unchanged, but at higher masses the trends of $\lsat$ flatten out due
to the larger cross sections of these halos.

The right-hand side of Figure \ref{lsat_theory} shows the same
results, but now binning by $\mgal$ rather than by $\mhalo$. As
expected by the scaling between $\mgal$ and $\mhalo$, the trend of
$\lsat$ with $\mgal$ is not as steep as before, but there is still a
clear power-law dependence of $\lsat\sim\mgal^{0.6}$ for
$\mgal<10^{10.5}$ $\msol$. At larger masses, the trend become much
steeper as the SHMR flattens out, meaning that a small change in
galaxy mass corresponds to a larger change in $\mhalo$. We note that
using fixed apertures does not change the scaling of $\lsat$ with
$\mgal$ for $\mgal<10^{10.5}$ $\msol$.

\begin{figure*}
\includegraphics[width=6in]{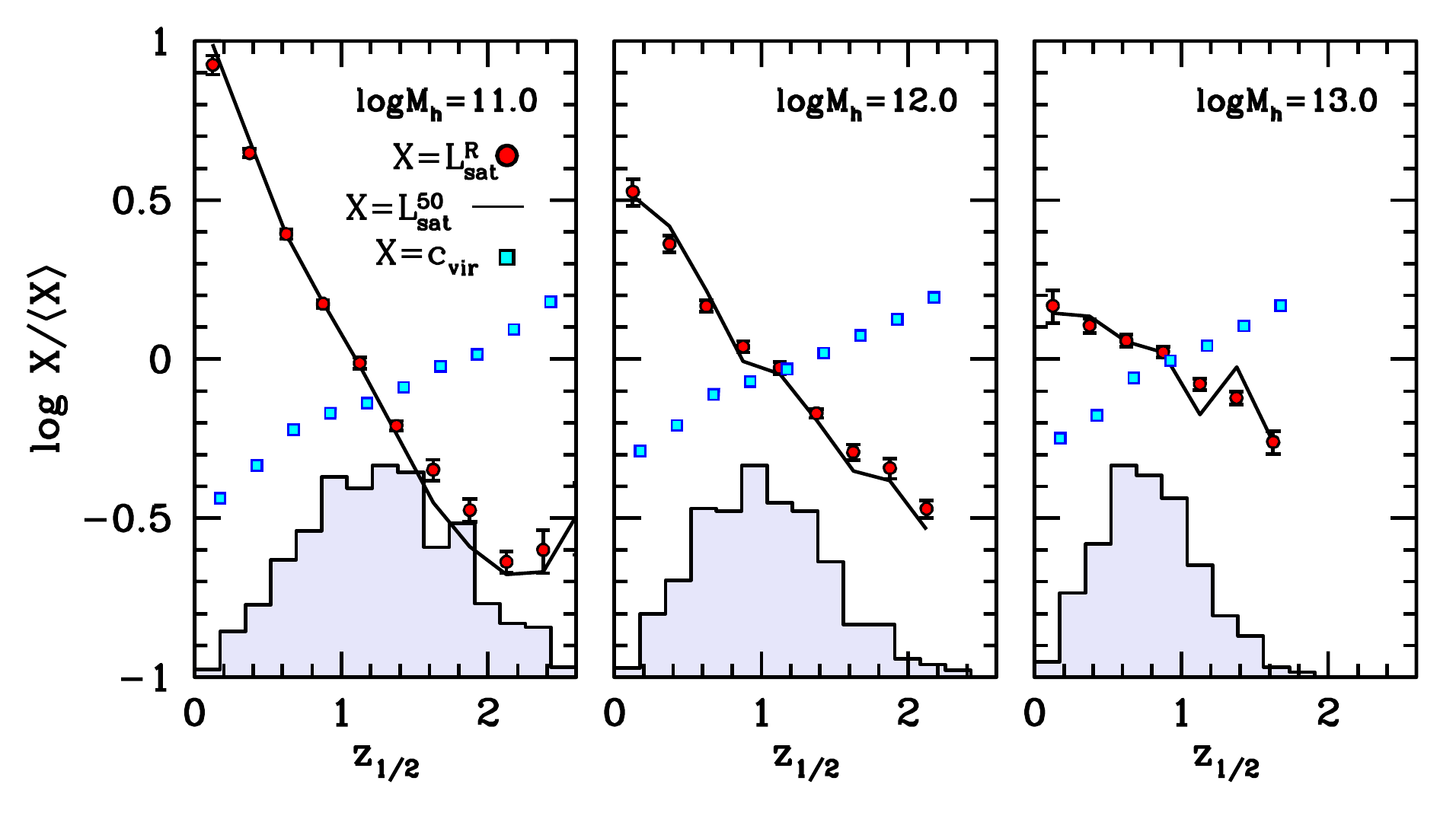}
\vspace{-0.5cm}
\caption{ \label{lsat_con3} Dependence of halo properties on formation
  time, for different halo masses. In each panel, red circles show the
  correlation between $\lsat(\rvir)$ and $\zhalf$ for halos of
  different masses. The distribution of $\zhalf$ is shown with the
  histogram at the bottom of each panel. Blue squares show the
  correlation between $c$ and $\zhalf$.  The solid curves show
  $\lsat(50)$, which is nearly the same as the results for
  $\rvir$. This indicates that concentration $c$ does not alter the
  relationship between $\zhalf$ and $\lsat$, regardless of the
  aperture used. All results are based on the C125 simulation. }
\end{figure*}

\subsection{Dependence of $\lsat$ on secondary halo properties}

Halos exhibit assembly bias (e.g., \citealt{wechsler_etal:06,
  gao_white:07}, and see \citealt{mao_etal:18, salcedo_etal:18,
  mansfield_kravtsov:19} for more recent treatments). At fixed mass,
certain secondary properties of a halo are correlated with its
large-scale environment, and thus impact halo clustering. One known
property that exhibits halo assembly bias is the amount of
substructure within the halo (\citealt{zentner_etal:05,
  wechsler_etal:06, gao_white:07, mao_etal:15}). Early-forming halos
have lower numbers of subhalos, $N_{\rm sub}$, because accreted
subhalos have had a longer amount of time to be tidally disrupted or
merge with the host halo due to dynamical friction. Late-forming halos
have had more recent accretion events, thus they will have a surplus
of substructure. Halo formation history correlates strongly with
large-scale environment, such that early-forming low-mass halos are in
higher-density regions. Here were parameterize halo formation time as
the redshift at which the halo reaches half its present-day mass,
$\zhalf$.

Another halo property that correlates with formation history is
concentration, $c$---early forming halos have higher
concentrations. For aperture measurements of $\lsat$ that are
significantly smaller than $\rvir$, we must determine the impact $c$
has on estimates of $\lsat$ at fixed halo mass.

Figure \ref{lsat_con3} shows the dependence of $\lsat$ on $\zhalf$ for
three different bins of $\mhalo$. The distribution of $\zhalf$ is
shown with the histogram at the bottom of each panel. The filled
circles show $\lsat(\rvir)$ while the solid curve shows
$\lsat(50)$. Both show a significant dependence of $\lsat$ on
$\zhalf$, with younger halos containing more satellites. Concentration
is highly correlated with $\zhalf$, as shown by the blue squares in
each panel. For these results, the $y$-axis is now
$\log c/\langle c\rangle$. Younger halos have lower concentrations,
while older halos have higher concentrations, in agreement with
previous results (e.g., \citealt{wechsler_etal:02,
  giocoli_etal:12}). The similarity between the circles and curves
shows that concentration does not alter the relationship between
$\zhalf$ and $\lsat$, even when only measuring $\lsat$ in the inner
parts of the halo.

\subsection{Dependence of $\lsat$ on large-scale environment}

\begin{figure}
  \includegraphics[width=3.5in]{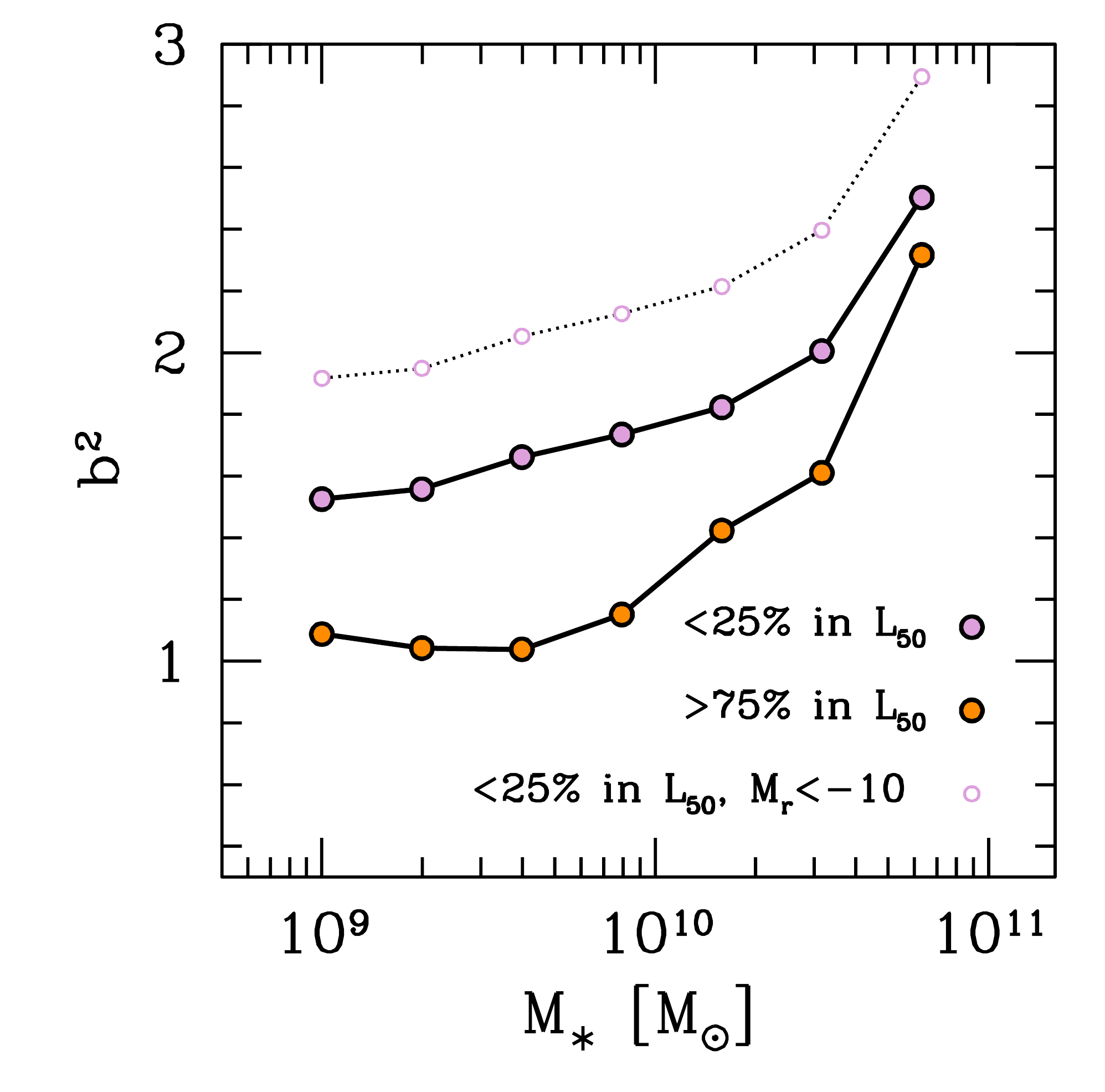}
  \vspace{-0.6cm}
  \caption{ \label{plot_bias} Clustering amplitude, expressed as
    $b^2$, as a function of $\mgal$. Results here are base on the C250
    simulation, combined with abundance matching models. The `bias
    factor' $b$ is defined as the ratio of the correlation function of
    halos to that of matter, $(\xi_h/\xi_m)^{1/2}$, at $r>10$
    $\hmpc$. Different colored circles indicate different quartiles in
    $\lsat$ at fixed $\mgal$. Central galaxies in halos with lower
    amounts of satellite light are more strongly clustered because
    these halos have early formation times. The difference between the
    clustering amplitude of the upper and lower quartiles is
    attenuated by the fact that many of the halos have no satellites
    about our fiducial magnitude threshold. Extending that threshold
    down to $M_r-5\log h=-10$ from $-14$ further separates the
    quartiles because a smaller fraction are empty. }
\end{figure}

The results of Figure \ref{lsat_con3} indicate that $\lsat$ will
correlate with large-scale density at fixed $\mhalo$ and $\mgal$. This
is shown explicitly in Figure \ref{plot_bias}, in which we plot the
clustering amplitude of galaxies relative to the clustering of dark
matter on large scales. Different curves show the top and bottom
quartiles of $\lsat$ at fixed $\mgal$. The lowest quartile in $\lsat$
has significantly higher clustering than the highest
quartile. However, the difference between the bias values is not as
large as it would be if the galaxies were partitioned by $\zhalf$
directly. This is due to the fact that a large fraction of halos,
especially at low $\mgal$, have no satellites brighter than
$M_r-5\log h=-14$. Thus, the trend of clustering with $\zhalf$ is lost
for the fainter part of the $\lsat$ distribution. For comparison, we
show the prediction for the bias of the faintest quartile assuming a
lower magnitude limit of -10 rather than -14. There are still many
empty halos, but the difference in clustering amplitudes is larger by
nearly a factor of two.

\subsection{Assembly bias and $\lsat$ scaling with $\mgal$}

Regardless of the degree to which $\lsat$ correlates with $\zhalf$, if
halo formation history does not correlate with galaxy properties then
it will not bias any results obtained through the $\lsat$
method---$\lsat$ would still be a true proxy for $\mhalo$ if primary
galaxies are stacked according to their observable properties. In
detail this is likely to be not exactly true---halo abundance matching
models based on $\vpeak$ or other formation-dependent halo quantities,
rather than halo mass, are a better match\footnote{We note that these
  conclusions are based on the standard $\chi^2$ statistic to
  discriminate between models. Other studies (\citealt{sinha_etal:18,
    vakili_etal:19}) have found that extra parameters above halo mass
  are not statistically preferred when using more sophisticated
  statistical analyses. Thus, the observational situation is not fully
  settled.} to observed galaxy clustering (\citealt{reddick_etal:13,
  lehmann_etal:17, zentner_etal:16}), and $\vpeak$ correlates weakly
with large-scale environment at fixed halo mass, amplifying halo
clustering by a few percent (\citealt{walsh_tinker:19}).

The right-hand panel of Figure \ref{lsat_theory} considers the
impact that galaxy assembly bias may have on how $\lsat$ scales with
$\mgal$. If, for example, $\mgal$ correlates with $\zhalf$ at fixed
$\mhalo$, then what is the impact on the $\lsat$-$\mgal$ correlation?
To test this idea, we use the conditional abundance matching framework
of \cite{hearin_watson:13}. After determining the mean
$\mgal$-$\mhalo$ relation through our parameterized SHMR, but before
adding scatter to the central galaxy $\mgal$, we bin all host halos in
narrow bins of $\mhalo$. In each bin, halos are rank-ordered by
$\zhalf$ and matched to the {\it residual} with respect to the mean
stellar mass, $\Delta\mgal$. The scatter about the mean is still a log-normal, but
now $\mgal$ is correlated with $\zhalf$ at fixed $\mhalo$. This yields
a galaxy-halo connection very similar to that seen in both
hydrodynamic models (\citealt{matthee_etal:17}) and semi-analytic
models (\citealt{tojeiro_etal:16}), in which older halos have more
massive galaxies at fixed $\mgal$. 

To estimate the maximal possible effect, we assume a 1:1 correlation
between $\zhalf$ and $\Delta\mgal$ with no scatter. The results of
this model for $\lsatx$ are shown in the right-hand panel of Figure
\ref{lsat_theory} with the orange cicles. Even with the maximal assembly
bias model, there is a negligible impact on $\lsat$ scaling. This
result would be quite different if we were binning on a secondary
galaxy property at fixed $\mgal$---properties such as effective radius
or galaxy velocity dispersion---and these properties correlated with
$\zhalf$. We will explore this possibility in a future work.

\subsection{Distinguishing $\mhalo$ from $\zhalf$ }

\begin{figure*}
  \includegraphics[width=6.5in]{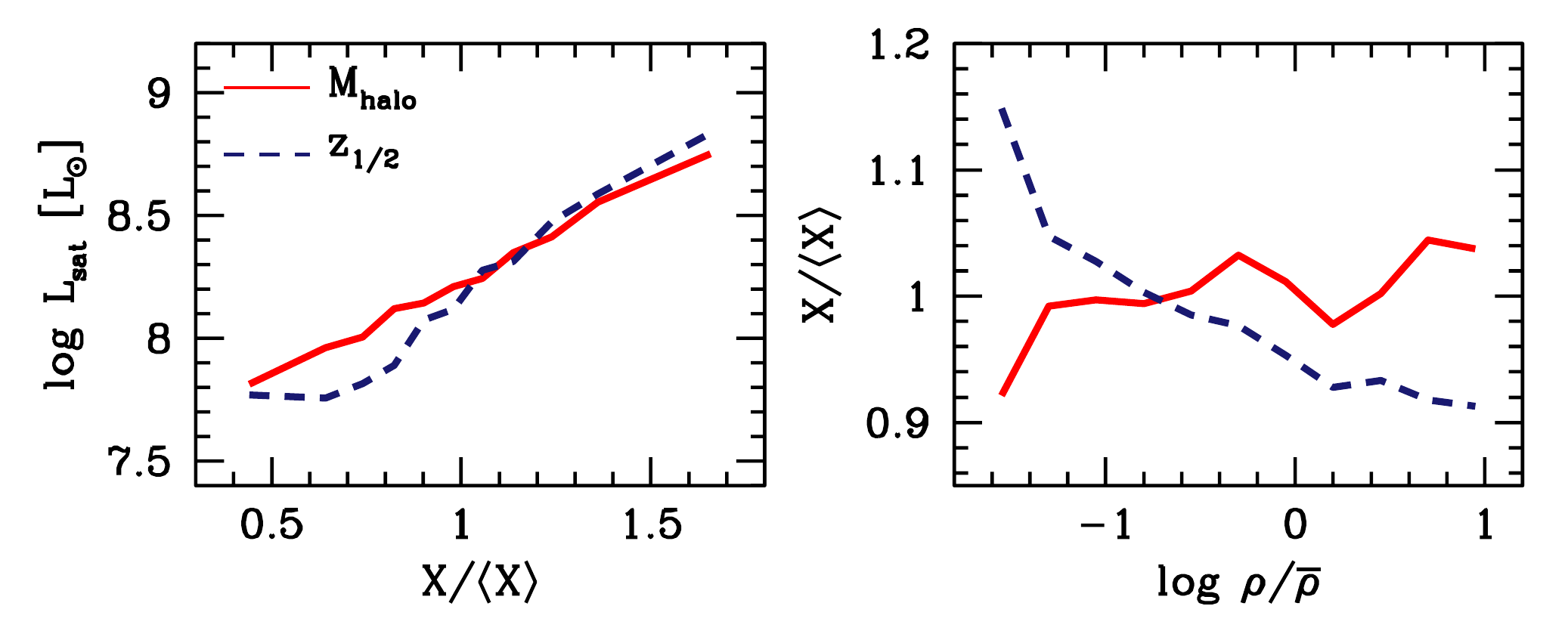}
  \vspace{-0.4cm}
  \caption{ \label{lsat_rho} {\it Left Panel:} Two theoretical models
    in which a hypothetical galaxy secondary property, $X$, correlates
    with the host halo at fixed $\mgal$. The solid red curve is a
    model in which $X$ correlates with $\mhalo$, yielding a
    correlation between $\lsat$ and $X$ at fixed stellar mass. The
    dashed blue line is a model in which $X$ anti-correlates with
    $\zhalf$ at fixed $\mgal$, yielding a similar correlation. These
    two models can be distinguished in the {\it right panel}, which
    shows the correlation between $X$ and large-scale density
    $\rho$. The $\zhalf$ model shows a correlation between $X$ and
    $\rho$, due to the fact that early-forming halos reside in
    low-density regions. In the $\mhalo$ model, the parameter $X$
    shows no correlation with density. All results are from the C250
    simulation.}
\end{figure*}

Figure \ref{lsat_rho} shows an example of how to distinguish between
the impact of $\mhalo$ and $\zhalf$ on $\lsat$. In the left panel, two
models are shown, constructed on the C250 simulation, in which $\lsat$
correlates with a hypothetical secondary galaxy, $X$, at fixed $\mgal$
for central galaxies. The galaxy stellar mass is $\sim 10^{10}$
$\msol$, so the halo mass is $\sim 10^{11.8}$ $\msol$.  For one model,
$X$ is correlated with $\mhalo$ at fixed $\mgal$, thus $\lsat$
increases with $X$, even though stellar mass is held constant. For the
second model, $X$ anti-correlates with $\zhalf$ at fixed
$\mgal$. Thus, higher values of property $X$ correlate with younger
halos, yielding a correlation between $X$ and $\lsat$ that is
consistent between the two models.

In the right-hand panel, we show the predictions these two models make
for correlations between $X$ and large-scale environment,
$\rho/\bar{\rho}$. Here, $\rho$ is measured in the same way as it
would in the redshift survey data---measuring the density in a galaxy
density field, in redshift space, in 10 Mpc top-hat spheres. This is
consistent with the approach to measuring large-sale environment taken
in \cite{tinker_etal:17_p1, tinker_etal:18_p3}. The model in which $X$
anti-correlates with $\zhalf$ is an example of galaxy assembly
bias. There is a clear anti-correlation between $X$ and
$\rho$---younger halos reside in underdensities, thus galaxies with
higher values of $X$ also are found in underdensities. The model in
which $X$ correlates with $\mhalo$ shows no correlation with
$\rho$---at the halo masses probed, there is little correlation
between $\mhalo$ and $\rho$. At higher values of $\mgal$, the host
halo masses will eventually get large enough such that a trend of $X$
with $\rho$ will be produced. But this trend will have opposite sign
to the $\zhalf$ model---a positive correlation between $X$ and
$\mhalo$ yields a positive correlation between $X$ and $\rho$ at
$\mhalo$ significantly larger than $10^{12}$ $\msol$.

Although the amplitude of the trend produced by the $\zhalf$ model is
only $\sim 20\%$, this amplitude is detectable in current SDSS
data. Trends of this amplitude have been detected in star formation
rate and Sersic index, while correlations with $\rho$ can be ruled out
in other parameters, such as galaxy color or quenched fraction
(\citealt{tinker_etal:17_p1, tinker_etal:18_p3}).

\section{Testing the Selection of Central Galaxies}

A critical aspect of this approach is the ability to select a pure
sample of central galaxies. In this paper we test two similar but
complementary approaches to identify central galaxies within a
spectroscopic galaxy sample: a galaxy group finder and a `central
galaxy finder,' as specified in \S \ref{s.data}. Although the halo
masses estimated by the group finder can be suspect, especially when
breaking the group sample into red and blue central galaxies, one of
the group finder's strengths is identifying which galaxies are central
and which are satellites. The code yields a quantity labeled $\psat$,
which roughly corresponds to the probability of a given galaxy being a
satellite in a larger halo. To characterize an entire galaxy
population, splitting at $\psat=0.5$ classifies all galaxies as one or
the other. This breakpoint, however, does introduce some impurities
into the sample of central galaxies of roughly 10-15\%
(\citealt{tinker_etal:11}). Defining a sample of `pure' central
galaxies, with $\psat<0.1$, removes the majority of impurities with
only a small decrease in the completeness of the sample. Defined in
this way, pure centrals are not biased in terms of the distribution of
environments in which central galaxies are found
(\citealt{tinker_etal:17_p1}).

The second method, the central galaxy finder, is more flexible but
less robust. We describe the method in Appendix A. Briefly, the method
is quite similar to the group finder, but makes no attempt to find
true groups. The algorithm first assigns halo masses to galaxy masses
using {\it inverse} abundance matching---in contrast to using
abundance matching to assign $\mgal$ to halos in simulations. We use
the tabulated stellar-to-halo mass relations of
\cite{behroozi_etal:13} rather than using actual galaxy counts. Thus,
the method does not require that a galaxy sample be volume-limited, or
contain a large number of galaxies. To classify whether a galaxy is a
central, all that is required is to know the masses of its nearest
neighbors. Using the properties of halos---their velocity dispersions
and density profiles---the code determines the probability that a
given galaxy is a central galaxy in its halo. We refer to this
quantity as $\psatiso$.

\subsection{Bias on the $\lsat$-$\mgal$ relation}
\label{s.lsat_test_centrals}
  
\begin{figure*}
\includegraphics[width=7in]{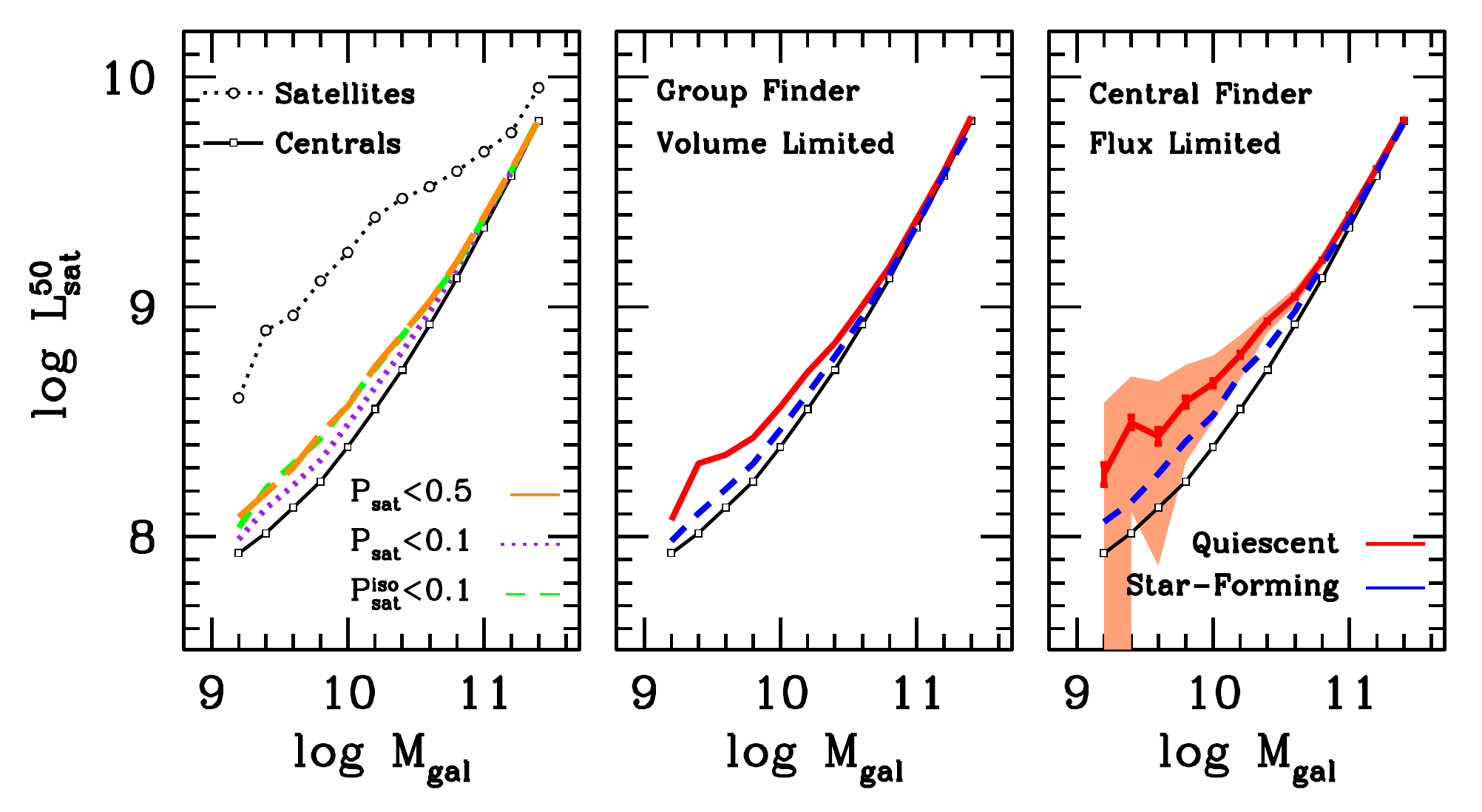}
\vspace{-0.7cm}
\caption{ \label{group_test} Impact of the selection of central
  galaxies of the values of $\lsatx$. The open circles connected by
  solid curve shows the input value of $\lsatx$ around central
  galaxies. The open circles connected by the dashed black curve shows
  the values of $\lsatx$ if satellite galaxies are taken to be the
  primary galaxies. Satellites reside in higher mass halos, thus their
  $\lsatx$ values are larger than those for central galaxies of the
  same $\mgal$. {\it Left hand panel:} $\lsatx$ values for the full
  sample of central galaxies constructed from mock galaxy
  distributions. The long-dashed and dotted curves show the results
  from applying the galaxy group finder to a volume-limited mock
  sample of galaxies. The long-dash line uses the full sample of
  central galaxies ($\psat<0.5$) while the dotted line shows results
  for the pure sample of central galaxies ($\psat<0.1$). The
  short-dashed line shows the results of applying the central galaxy
  finder to a flux-limited mock galaxy catalog, using $\pcen>0.9$ to
  define the sample. {\it Middle Panel:} $\lsatx$ values obtained from
  applying the group finder to a volume-limited mock galaxy sample,
  now broken into central galaxies that are star-forming and
  quiescent. See text for details. The central galaxy sample is
  defined as $\psat<0.1$. The input model assumes that all galaxies at
  fixed $\mgal$ live in the same mass halos, so the any difference in
  the measured $\lsatx$ values represents bias incurred by the group
  finding algorithm. {\it Right Panel:} Same as the middle panel, only
  now the mock galaxy sample is flux-limited, matching the $n(z)$ of
  the SDSS data, and central galaxies are identified using the central
  galaxy finder with $\psatiso>0.9$. The errors on the results for
  quiescent galaxies represent the error in the mean of 100
  realizations of the mock. The shaded area represents the
  dispersion.}
\end{figure*}

Figure \ref{group_test} shows the impact of impurities in the central
galaxy samples has on $\lsatx$. To perform this test, we created mock
galaxy distributions using the C250 box that match the PCA stellar
mass function. Each galaxy in the mock is assigned both an $\lsatx$
value and total luminosity of interloper, or `background' galaxies we
term $\lbg$. Thus the total light within the aperture is
$\ltot=\lsatx + \lbg$. These are chosen to match the distributions of
$\ltot$ and $\lbg$ seen in the data, which we describe in \S
\ref{s.application}. Results in Figure \ref{group_test} represent the
mean of 100 mock realizations. In each realization, the quantity
that varies is the assignment of $\lsatx$, which is drawn randomly
from the distribution of $\ltot$ and $\lbg$ in the SDSS data.

Central and satellite galaxies follow different distributions of
$\lsatx$. The left-side panel of Figure \ref{group_test} shows the
input values of $\lsatx$ as a function of $\mgal$ of the mock. At a
given $\mgal$, the $\lsat$ values for satellite galaxies is roughly an
order of magnitude higher than for centrals. Thus, impurities in the
central sample can bias the measurements if the purity goes
significantly below unity. 

To construct a mock galaxy survey suitable for testing the group
finder, we first convert the cartesian positions and velocities of the
cubic mock from C250 to RA, Dec, and redshift assuming one corner of
the box as the observer. The resulting mock is an octant of the full
sky, volume-limited down to $\mgal=10^9$ $\msol$, with a maximum
redshift of $z=0.08$. The group finder is then applied to the mock. We
use the group finder results to make two central galaxy samples: one
with $\psat<0.5$ and a pure sample with $\psat<0.1$. The results are
shown with the long-dash and dotted lines in the left-hand
panel. Above $\mgal=10^{10.5}$ $\msol$, the bias induced by the
impurities in the sample are negligible. At lower masses, the
$\psat<0.5$ sample yields a bias of $\sim 0.15$ dex. The pure sample
of $\psat<0.1$ yields a bias roughly half as large, at $\sim 0.08$
dex.

To apply the central galaxy finder to the mock, we modify the
procedure to construct a flux-limited sample, rather than a
volume-limited sample. First, we repeat the C250 cubic mock 8 times to
extend the maximum redshift limit to $z=0.16$. The mock is divided
into narrow bins of $\log \mgal$, which are subsampled match the
observed redshift distributions $N(z|\mgal)$ in the full flux-limited
SDSS catalog, in units of [dz$^{-1}$deg$^{-2}$.] We do this process
separately for star-forming and quiescent galaxies, because quiescent
galaxies are fainter at fixed $\mgal$, thus they have distinct
redshift distributions.

We apply the central galaxy finder to the flux-limited mock catalog,
using $\psatiso>0.9$ to define the sample of central galaxies. The
$\lsatx$ results, shown in the left-hand panel of Figure
\ref{group_test} as well, are nearly indistinguishable from the full
$\psat<0.5$ sample from the group finder. Thus, the central galaxy
sample from the flux-limited catalog has slightly more bias in it than
the pure sample from the volume-limited catalog, but the flexibility
increased statistics afforded through the flux-limited catalog make it
possible to perform fine-grained, multi-dimensional binning on $\lsat$
results. 

\subsection{Bias on $\lsat$ in sub-populations} 

The true concern of impurities in the central sample is not in biasing
the results of $\lsatx$ for the full sample, but rather creating a
differential bias for sub-classes of galaxies that have a higher
fraction of satellite galaxies in them. The prime example of such a
sample is quiescent galaxies---quiescent galaxies have a significantly
higher satellite fraction than star-formning galaxies at all stellar
masses (e.g., \citealt{weinmann_etal:06, wetzel_etal:12,
  tinker_etal:13}). In the mock, we constrain the quenched fraction to
match that seen in the group finder as a function of stellar mass. We
do this separately for true centrals and true satellites in the mock,
as they have different quenched fractions. We assume that star-forming
and quiescent central galaxies of the same $\mgal$ live in halos of
the same $\mhalo$. This choice is driven less by results in the
literature (see the wide disparity of results in
\citealt{wechsler_tinker:18} described in \S 1) than it is to make it
straightforward to identify any biases imparted by impurities in the
central sample: the input model yields $\lsatx$-$\mgal$ correlations
that are identical for star-forming and quiescent galaxies. Thus any
differences in the final results are a consequence of impurities in
the central galaxy sample.

The middle panel of Figure \ref{group_test} shows the results of
applying the group finder to the volume-limited mock, and then
dividing up the sample by star-forming and quiescent galaxies. Results
here are for $\psat<0.1$. The results for star-forming galaxies are
nearly identical to those for the full sample in the left-hand panel,
but the quiescent galaxies have slightly higher $\lsatx$ values,
roughly twice the bias (in dex). 

The right-hand panel of Figure \ref{group_test} shows the results of
applying the central galaxy finder to the flux-limited mock
catalog. The amplitudes of the biases are larger than that seen in the
results from the group catalog, but the results are consistent, with
the bias in $\log\lsatx$ for quiescent galaxies being twice as large
as that seen in the star-forming central galaxies. 

Although the biases seen in Figure \ref{group_test} are
non-negligible, they are small compared to the overall trend in
$\lsatx$ with $\mgal$, and can be corrected using the results of mock
tests such as this one.

\begin{figure}
\includegraphics[width=3.3in]{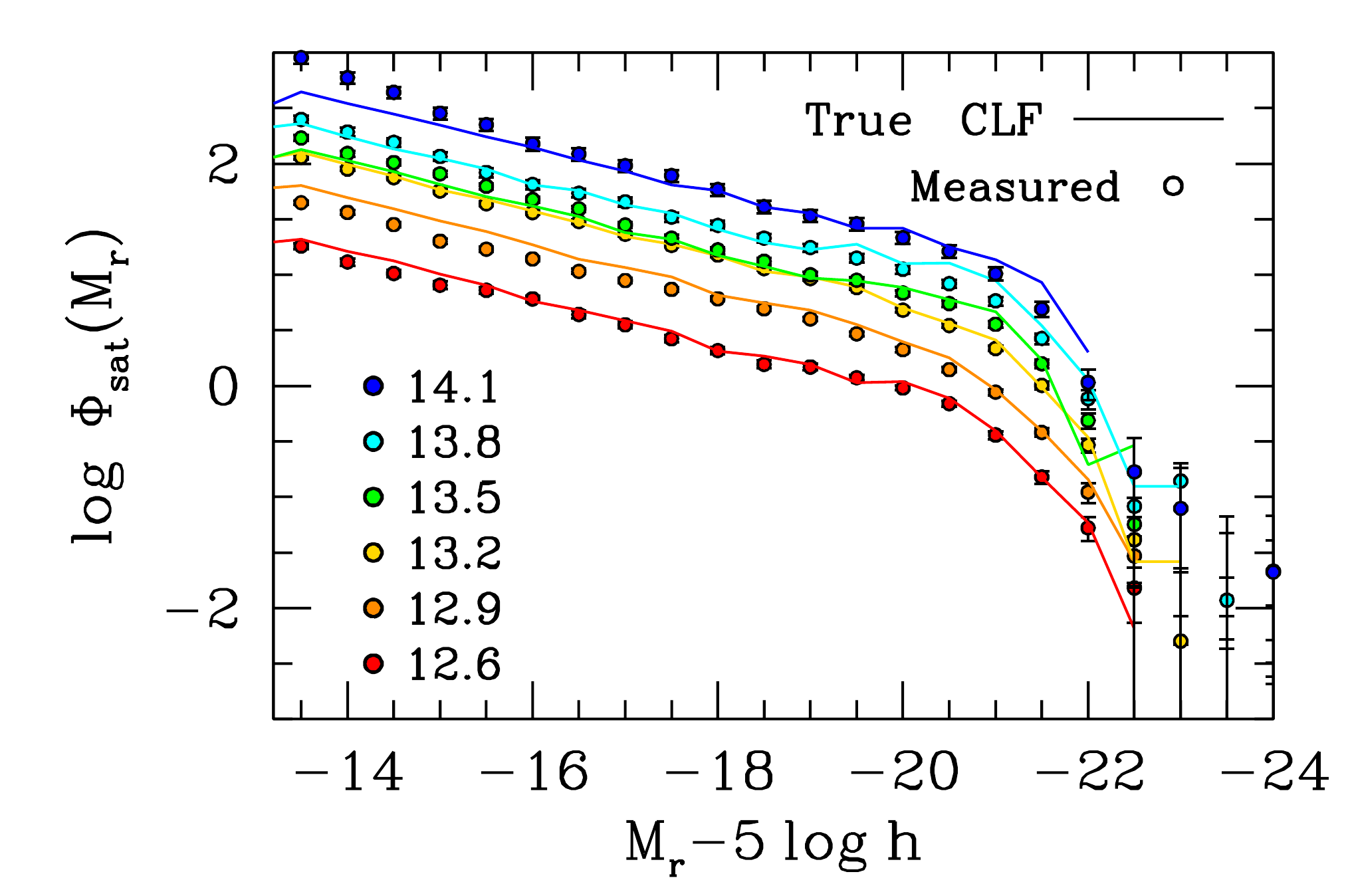}
\caption{ \label{mock_lumfunc} Recovery of conditional luminosity
  functions in simulations. Here we present results using the Buzzard
  mock galaxy samples (\citealt{buzzard}). The Buzzard simulations are
  constructed to match observational statistics of an $r<24$
  photometric sample of galaxies (citation). The method of measuring
  $\phisat$ uses annuli around each primary to estimate the
  background. We find minimal dependence of results on the exact
  choice of the annuli radii, with the exception of the highest mass
  bin, $\mhalo>10^{14}$ $\msol$. }
\end{figure}

\begin{figure}
\includegraphics[width=3.in]{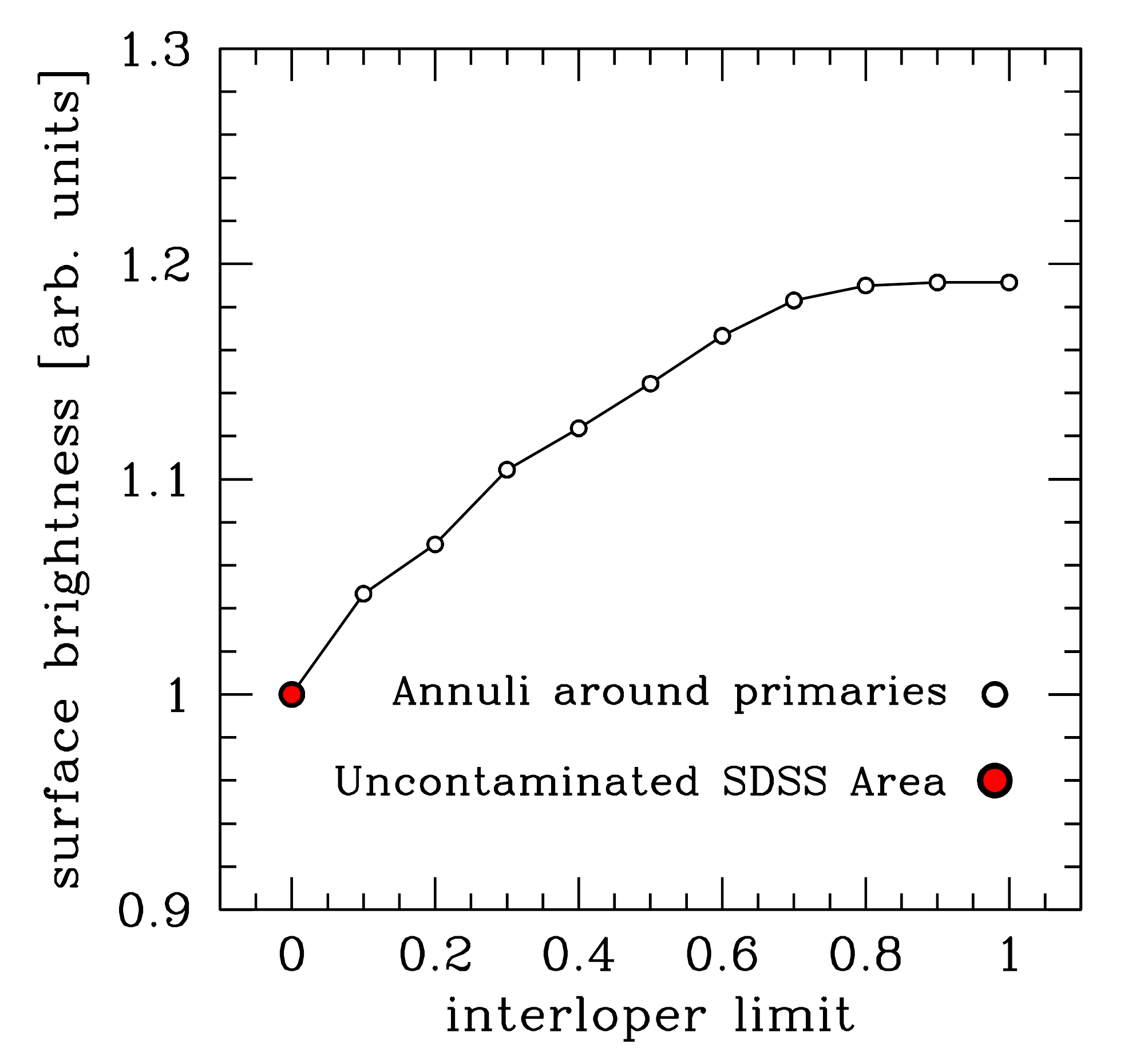}
\caption{ \label{systest_background} Open circles show the estimates of
  the surface brightness in background galaxies using annuli around
  SDSS primary galaxies. All annuli have some contamination by interlopers
  (other primary galaxies). The annuli are rank-ordere by total stellar
  mass of SDSS galaxies within them, and thus `interloper limit'
  indicates what fraction of annuli are used to calculate the
  background. The filled red circles at $x=0$ is an estimate of the
  background from random locations that have no interlopers. }
\end{figure}

\begin{figure*}
\includegraphics[width=6in]{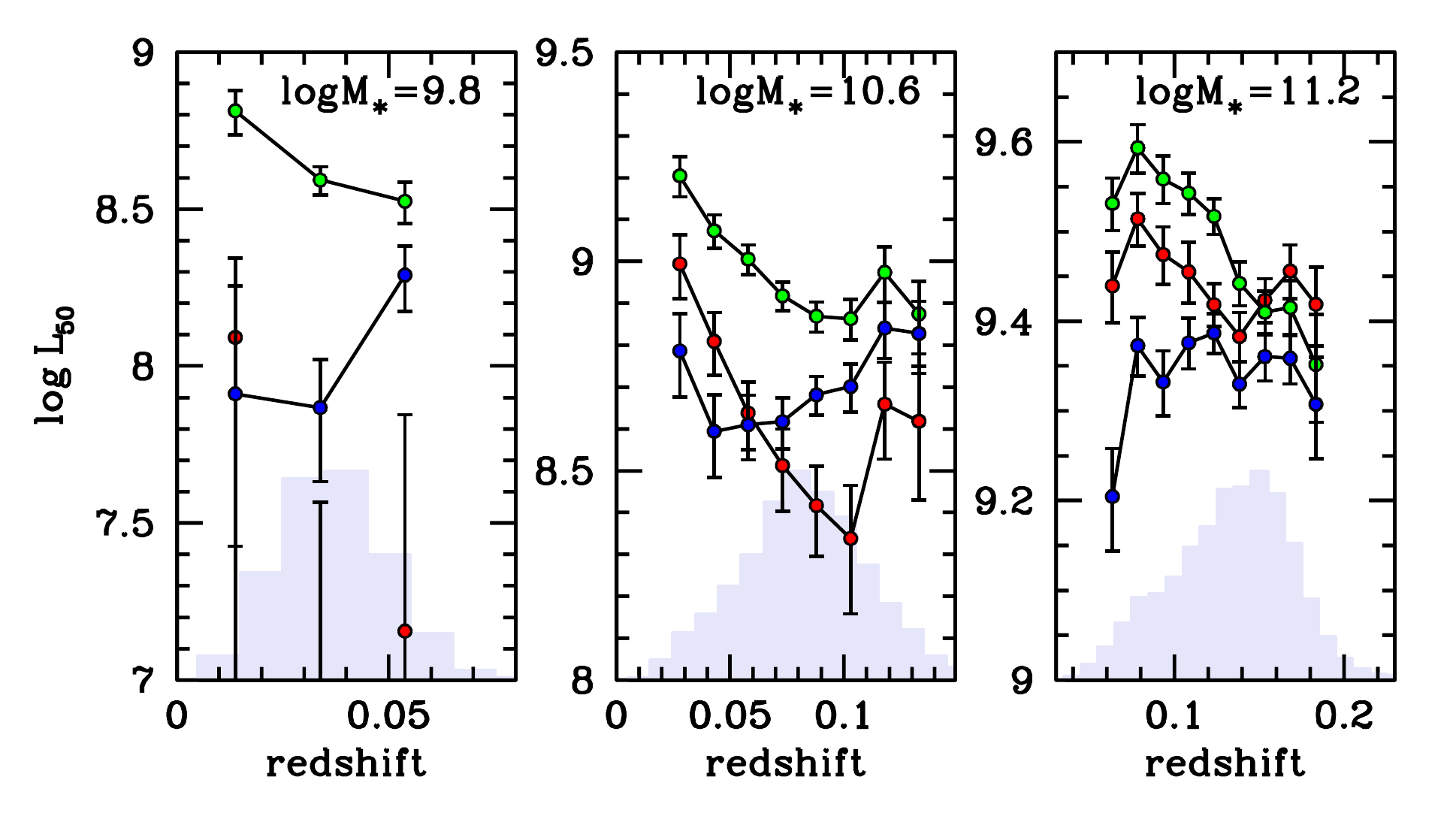}
\vspace{-0.4cm}
\caption{ \label{systest_redshift} Results for SDSS primary galaxies:
  $\lsatx$ as a function of $z$ for three different methods of
  measuring $\lsat$. In each panel, the blue points indicate $\lsatx$
  after removing primary galaxies that have interlopers inside the 50
  $\hkpc$ aperture, and the background luminosity of galaxies is taken
  from random points within the SDSS footprint (see text for
  details). This method produces $lsatx$ results that are roughly
  constant with redshift, and thus we use it as our fiducial method.
  Red points indicate $\lsatx$ when using annuli around the primaries
  to estimate the background, and including all primaries that are
  within the DLIS footprint. This method yields a strong redshift
  trend, and even produces some negative values of $\lsatx$.  Green
  points represent $\lsatx$ when using random points to estimate the
  background, but still include all primary galaxies. This method also
  produces a strong redshift trend. Results are shown for three
  different bins in $\mgal$. Below $\mgal=10^{9.8}$ $\msol$, the data
  do not span a wide enough range in redshift to quantify and trend in
  $z$. The shaded histogram at the bottom of each panel shows the
  redshift distribution for galaxies in the $\mgal$ bin. }
\end{figure*}

\section{Application to Data}
\label{s.application}

In this section we will discuss how we ameliorate observational
systematics that may influence the measured values of $\lsatr$ and
$\lsatx$. We focus here on variable survey depth for $\lsatr$,
unbiased estimation of the background galaxy counts for $\lsatx$, and
eliminating redshift dependence of the $\lsatx$ measurements. The
results presented in this section will show measurements conducted on
actual Legacy Surveys data.

\subsection{Measuring the conditional luminosity function within dark
  matter halos}

The conditional luminosity function (CLF) is defined as the luminosity
function of satellite galaxies, conditioned on the mass of the host
halo, $\phisat(M_r|\mhalo)$.  The depth of the DLIS data varies across
the sky, thus when measuring the CLFs, we track the limiting magnitude
for each halo depending on its angular position and redshift. For each
primary galaxy, we assume all photometric galaxies along the line of
sight are at the redshift of the primary galaxy.  For each halo, we
define the limiting absolute magnitude
$M_{r,\rm lim}=m_{r,\rm lim}+\mu$, where $\mu$ is the distance modulus
at the redshift of that halo and $m_{r, \rm lim}$ is the limiting
magnitude at that location in the survey.  The luminosity
function is defined as

\begin{equation}
\label{e.phi}
\phisat(M_r|\mhalo)dM_r = \frac{N_{\rm gal}(M_r)}{N_h(M_r)},
\end{equation}

\noindent where $N_h(M_r)$ is the number of halos that have a
$M_{r,\rm lim} >M_r$, and $N_{\rm gal}(M_r)$ is the number of galaxies
above the background at magnitude $M_r$, expressed as

\begin{equation}
  \label{e.ngal}
N_{\rm gal}(M_r) = N_{\rm tot}(M_r) - f_A N_{\rm BG}(M_r),
\end{equation}

\noindent where $N_{\rm tot}$ is the total number of galaxies at $M_r$
within the aperture centered on the SDSS galaxy, and $N_{\rm BG}$ is
the number of background galaxies at that magnitude. This quantity is
measured from an annulus around the halo. This is the same approach
as taken in \cite{hansen_etal:09} and \cite{tal_etal:12} for
estimating the background around massive objects, such as clusters or
luminous red galaxies. Our fiducial choice for the annuli radii are
$R=[\rvir,3\rvir]$, but in practice we find that the exact choice of
annuli boundaries has negligible impact on the results. The factor
$f_A$ is to account for any differences in the area used to estimate
the two $N$ values. For example, if we use annuli with
$R=[\rvir,3\rvir]$, $f_A$ for $\lsatr$ is $1/(9-1)=1/8$.

\subsection{CLF recovery in simulations}

Figure \ref{mock_lumfunc} shows the results of applying our method of
measuring the CLF on simulations. The mock galaxy catalogs we use are
the Buzzard mocks (\citealt{buzzard}). These simulations were utilized
by the Dark Energy Survey (DES) to test the pipeline and quantify
systematic errors in the cosmological analysis
(\citealt{maccrann_etal:18}). The Buzzard mock galaxy distributions
have large volume, subtending roughly 10,000 deg$^2$ when projected
onto the sky. The mocks also incorporate galaxies faint enough to
match the flux limit in DES imaging, which is significantly deeper
than DLIS data. The mocks are tuned to match the evolution of the
luminosity function from $z=0$ to $z=1$, galaxy clustering as a
function of luminosity, and the observed color-density relation at $z=0$. In
order to create such a large-volume mock, the dark matter simulation
on which the mock is built only resolves halos down to
$\mhalo\approx 10^{12.5}$ $\msol$, and the galaxies that would be
contained in lower-mass halos are placed by sampling the dark matter
density field in order match all of the observational statistics
mentioned above.

To mimic our analysis of $\lsatr$ on DLIS data, we restrict the Buzzard
mock to galaxies with $r<24$, and construct a volume-limited sample of
primary galaxies from all resolved halos within $z=0.1$. Here we
assume perfect knowledge of the true sample of central galaxies, thus
this test isolates the method of stacking halos and background
subtraction. As with the DLIS data, we measure the background using
annuli around each primary object. Results are shown in Figure
\ref{mock_lumfunc} in bins of $\log\mhalo$. The recovered CLFs are in
excellent agreement with the mock inputs, including how the amplitude
of $\phisat$ scales with $\mhalo$ and the magnitude of the cutoff in
each luminosity function. This test demonstrates that the interlopers
present in both the annuli and the apertures around the primary
galaxies cancel each other, allowing robust recovery of $\phisat$. To
measure $\lsatr$, we integrate the measured values of
$\phisat(M_r|\mhalo)$, weighted by the luminosity at each bin of
$M_r$.

\subsection{Measuring $\lsatx$}

As discussed above, when measuring $\lsat$ around galaxies---rather
than halos---we wish to impart no prior on $M_h$ in the
measurements. Thus we choose $\lsatx$ as our observational quantity
around galaxies.  To measure $\lsatx$ we first estimate the CLF
individually for each galaxy, but restricting the galaxy counts to be
within the 50 $\hkpc$ aperture centered on the primary galaxy. Our
method for measuring $\lsatx$ differs from $\lsatr$ in out approach to
estimating $N_{\rm BG}(M_r)$ in equation (\ref{e.ngal}), which we we
discuss in the proceeding section.  After making the measurement of
the 50-$\hkpc$ CLF, we integrate $\Phi(M_r)$ from $M_r^{lim}$:

\begin{equation}
  \label{e.lsat}
\lsat = \sum_{M_r^{\rm low}}^{M_r^{\rm hi}} 10^{-0.4(M_r-M_{r,\odot})} \Phi(M_r)\Delta M_r,
\end{equation}

\noindent where $M_{r,\odot}=4.65$ and we use the center of the bin
for the value of $M_r$. In Eq. (\ref{e.lsat}), the limits of the
summation are set to maximize the signal to noise ot the $\lsat$
measurement. In practice, background galaxies at low redshift (but
below the magnitude limit of SDSS spectroscopy) can cause large
fluctuations in $\lsat$ when converting to absolute magnitude. Thus we
enforce a bright limit based on the stellar mass of the central
galaxy, such that

\begin{equation}
  \label{e.maghi}
  M_r^{\rm hi} = -21 - 2\times (\log\mgal-10).
\end{equation}

\noindent Essentially, Eq. (\ref{e.maghi}) enforces a limit that
satellite galaxies cannot be brighter than the central galaxy
itself. For $M_r^{\rm low}$ we chose $M_r-5\log h=-14$. At fainter
magnitudes, we find that the luminosity function of all galaxies
within the aperture centered on the central galaxy falls {\it below}
the luminosity function of the background at fainter magnitudes (i.e.,
equation (\ref{e.ngal}) would be negative). The presence of bright
foreground galaxies---i.e., the SDSS centrals and any satellites
within the halo---reduces the efficiency of finding faint extended
objects in the DLIS imagery. We choose -14 as our faint limit
because it yields a nearly volume-limited sample out to $z=0.15$,
which is the maximum redshift in our flux-limited sample.

\subsection{Estimating the background for $\lsatx$}

As mentioned above, we use a different approach to estimating the
background surface density of galaxies when measuring $\lsatx$ than
when measuring satellite luminosities within the entire halo. When
using annuli at $\sim 3\rvir$ to estimate the background, most of
these annuli are impacted by the presence of interlopers---other SDSS
galaxies---projected along the line of sight. Thus, the estimate of of
$\lbg$ is biased due to the presence of halos that contain an enhanced
number of secondary galaxies within them. 

When estimating $\lsatr$, using annuli does not actually bias the
value of $\lsat$ obtained---the apertures subtended by the halo
$\rvir$ suffers from the same contamination at the annuli themselves
(as demonstrated in the comparison to the Buzzard mocks in Figure
\ref{mock_lumfunc}). However, when estimating $\lsatx$, the smaller
aperture yields a much smaller fraction of $\ltot$ measurements that
have interlopers. Although it is possible to develop an estimator to
robustly account for the presence of interlopers in the background
(i.e., \citealt{masjedi_etal:06}), these primary objects can be
successfully removed from the sample of central galaxies without
significantly impacting the statistics of the overall
sample. Therefore, to measure an unbiased $\lsatx$ we require an
estimate of $\lbg$ that is also unaffected by interlopers.

Figure \ref{systest_background} shows how interlopers impact the value
of $\lbg$. The $y$-axis shows the mean surface brightness of
background galaxies in annuli around primary galaxies with
$\mgal=10^{10.5}$ $\msol$. All these annuli contain interlopers within
them, thus for each annulus we measure the total stellar mass of all
interlopers within the annulus. The $x$-axis, labeled `interloper
limit,' refers to the the fraction of annuli used in calculating the
mean surface brightness of background galaxies. The annuli are
rank-ordered by the total stellar mass of the interlopers. An
interloper limit of 0.6 means than the lower 60\% of annuli are used
to calculate the mean background surface brightness. Thus, the
background surface brightness decreases monotonically with decreasing
interloper limit.

The red circle at zero on the $x$-axis is an estimate of the
background from randomly-chosen areas within the footprint that do not
contain interlopers. Specifically, this value is the average of
$\sim 30,000$ apertures with angular radius 35.2 arcsec, which is 50
$\hkpc$ at $z=0.05$. The locations of these points were chosen from
random locations within both the SDSS mask and the DLIS footprint, and
then restricted to only include locations that are outside $\rvir/2$
of the nearest SDSS galaxy. Extending this limit to only include
locations that are fully outside any SDSS virial radius does not leave
enough area for a robust estimation of the background. But, as can be
seen from Figure \ref{systest_background}, using these random
locations yields an estimate of the background that is a clear
extrapolation of the annuli toward having no interlopers. In all
measurements of $\lsatx$, we use the the mean
$\langle N_{\rm BG}(M_r)\rangle$ from these 30,000 apertures as our
estimate of the background surface density of galaxies.

\subsection{Redshift effects on $\lsatx$}

We do not expect the galaxy-halo connection to change significantly
over the redshift baseline of the SDSS. Thus, a good diagnostic for
our method of estimating $\lsat$ is to demonstrate that the quantity
is independent of $z$. Figure \ref{systest_redshift} shows results
with DLIS data to demonstrate this fact. 

Figure \ref{systest_redshift} shows $\lsatx$ as a function of $z$ for
three different stellar mass bins. In each panel, we show three
different methods of calculating $\lsatx$: Red points show
measurements using annuli to estimate the background, and all primary
galaxies are included regardless of whether there are interlopers
within the 50 $\hkpc$ aperture. Green points show the results for all
primary galaxies when we use the uncontaminated random points for the
estimate of $\lbg$. Blue points---our fiducial method---show the
results using the random points for $\lbg$, but now removing primary
objects with interlopers within the aperture. The shaded histogram at
the bottom of each panel shows the redshift distribution for each
sample of galaxies.

In each panel, the red points (annuli background) show significant
redshift trend of $\lsatx$. For the lowest $\mgal$ bin, the values of
$\lsatx$ become negative at the peak of the redshift
distribution. Switching from annuli backgrounds to the random
background avoids this problem, since now the estimate of $\lbg$ is
$\sim 20\%$ lower, but there is still a significant redshift trend
with $\lsatx$. The blue points, where primary objects with interlopers
are removed, yields $\lsatx$ values that are roughly independent of
redshift. Errors on each data point are from bootstrap resampling on
the sample of primary galaxies in each bin in redshift.


\begin{figure*}
\includegraphics[width=7in ]{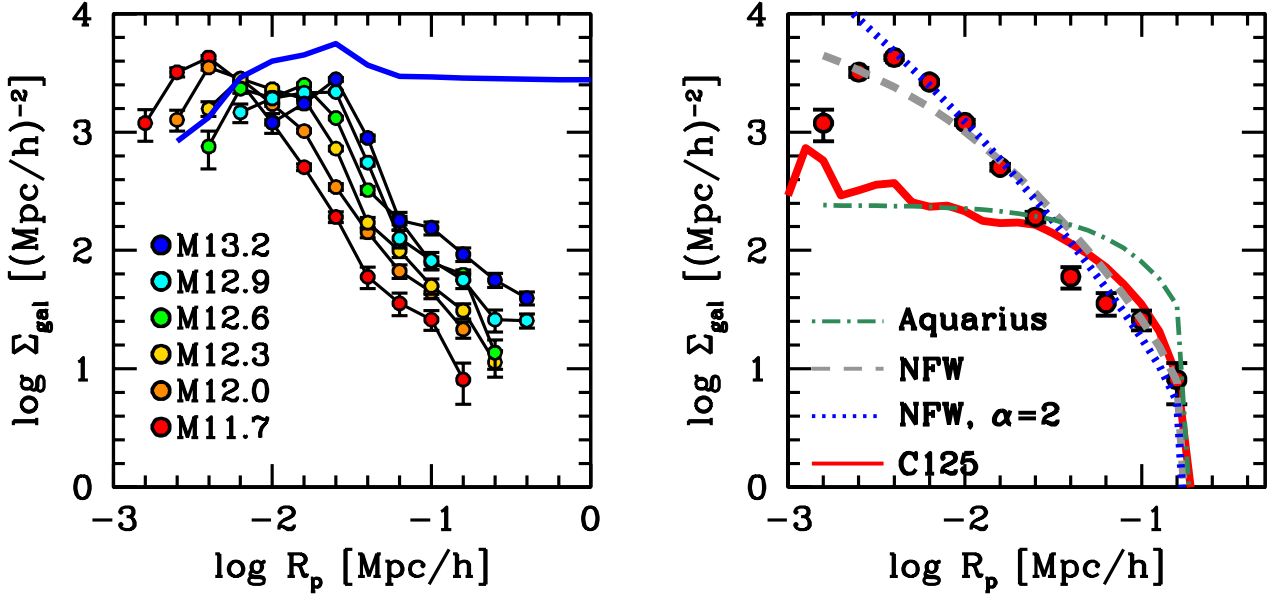}
\vspace{-0.8cm}
\caption{ \label{radial_profile} {\it Left Panel:} The excess surface
  density of galaxies above random, $\Sigma_{\rm gal}$, as a function
  of projected comoving separation from the group center. To calculate
  distances, all galaxies are assumed to be at the redshift of the
  group. The key shows the value of $\log\mhalo$. The solid blue curve
  shows the total projected number of densities of galaxies before
  background subtraction. All data are for galaxies brighter than
  $M_r-5\log h=-14$. The turnover of all these data at small scales is
  due to the presence of a bright central galaxy in the group. {\it
    Right Panel:} $\Sigma_{\rm gal}$ for the $\mhalo=10^{11.7}$ bin
  compared to different theoretical models. The solid red curve shows
  the abundance matching result from the C125 simulation, which is
  consistent with the higher-resolution Aquarius simulations. The dash
  and dotted curves show NFW fits with concentration parameter $c=15$,
  but with inner slopes of $\gamma=-1$ (standard NFW) and $\gamma=-2$,
  which is supported by small-scale clustering measurements of
  \citet{watson_etal:12}.  }
\end{figure*}

\section{Results for SDSS Central Galaxies}
\label{s.results}

In this section, we present our first results of measuring $\lsat$
around SDSS galaxies. We focus on how our measurements of $\lsat$
scale with $\mhalo$ and $\mgal$, and how they compare to our abundance
matching predictions. The values of $\mhalo$ are taken from the group
catalog, and thus should be taken as estimates only. Analysis of the
$\lsat$ results for sub-populations of galaxies, secondary galaxy
properties, and correlations with environment will be presented in
further work.

\subsection{Projected Radial Profiles}

Figure \ref{radial_profile} shows the surface number density of
photometric galaxies, centered on primary identified through
the galaxy group finder. In the left-hand panel, each color indicates a
different bin in $\mhalo$. As halo mass increases, the amplitude of
$\Sigma_{\rm gal}$ also increases, but as $R_p$ decreases, each
projected density profile hits a maximum value and then turns over at
smaller scales. This is due to the presence of a bright central galaxy
located at the center of the halo. The location of this peak value
increases with $\mhalo$, as expected as the luminosity of the central
galaxy increases as well. The thick blue curve in this panel shows the
total projected number density of galaxies before background
subtraction, for $\mhalo = 10^{13.2}$ $\msol$. The value of the
background can easily be seen, as $\Sigma_{\rm gal}$ reaches a
horizontal asymptot well within the virial radius of the halo.
However, this curve shows the same behavior as the measurements for
satellite galaxies, peaking at $R_p\sim 15$ kpc/h. Inside this scale,
the total $\Sigma_{\rm gal}$ falls below the asymptotic value of the
background, demonstrating the presence of the central galaxy inhibits
source detection at the center of the halo.

The right-hand panel of Figure \ref{radial_profile} compares our
measurements for $\mhalo=10^{11.1}$ $\msol$ halos to several
theoretical predictions. The solid red curve shows the abundance
matching prediction from the C125 simulation. The thick dashed line
show NFW profiles (\citealt{nfw:97}) with concentration parameter of
$c=15$, which is consistent with the $\lcdm$ prediction for halos of
this mass (\citealt{maccio_etal:08}). The blue dotted line is a
modified NFW profile, with inner slope $\gamma=-2$ rather than -1, but
with the same concentration parameter. The dot-dash line shows the
Einasto profile fit to the distribution of subhalos in the Aquarius
project (\citealt{aquarius}). 

The abundance matching prediction is significantly shallower than the
SDSS results or the NFW profiles, although it is in good agreement
with the Aquarius results, which are significantly higher
resolution. Whether these simulation results are still affected by
numerical artifacts is still an open issue however;
\cite{vandenbosch_etal:18b, vandenbosch_etal:18} find that subhalo
disruption via physical mechanisms should be occur only very rarely,
and thus the disruption of substructure seen in N-body simulations is
largely artificial. There is observational evidence consistent with
this claim as well: when constructing models of faint Milky Way
satellite from collisionless N-body simulations, \cite{nadler_etal:19}
note that ``orphans''---evolving satellite galaxies analytically after
their host subhalos are too disrupted to track within the
simulation---are required to match the total observed number of
satellites. Previous comparisons between the small-scale clustering of
subhalos and measured clustering of galaxies have shown good agreement
(e.g., \citealt{conroy_etal:06, reddick_etal:13, lehmann_etal:17}),
but these comparisons are restricted to scales larger than those of
interest here ($R_p\ga 0.2$ $\hmpc$) and for brighter galaxy
samples. Clustering measurements at these scales are largely
insensitive to the values of halo concentration parameters used
(\citealt{tinker_etal:12_mn}). \cite{watson_etal:12} use
cross-correlation techniques to measure galaxy clustering down to
$R_p\sim 0.01$ $\hmpc$, finding evidence for steeper inner density
profiles, consistent with the $\gamma=-2$ shown in Figure
\ref{radial_profile}, and concentration parameters roughly consistent
with that predicted by collisionless N-body simulations.

Either NFW profile is a reasonable description of the data, relative
to the subhalo results, but all $\Sigma_{\rm gal}$ measurements have a
``kink'' at $R_p\sim \rvir/5$. This feature can be seen in the results
for the lowest halo mass bin in the right-hand panel of Figure
\ref{radial_profile}, at $\log R_p\approx -1.4$, where the SDSS
measurements fall below the predictions for C125. This could be an
artifact of the small amount of satellite galaxies leaking into the
sample of central galaxies, or an observational systematic related to
background subtraction. Further study is required for a sufficient
explanation of this feature.

\begin{figure*}
\includegraphics[width=7in ]{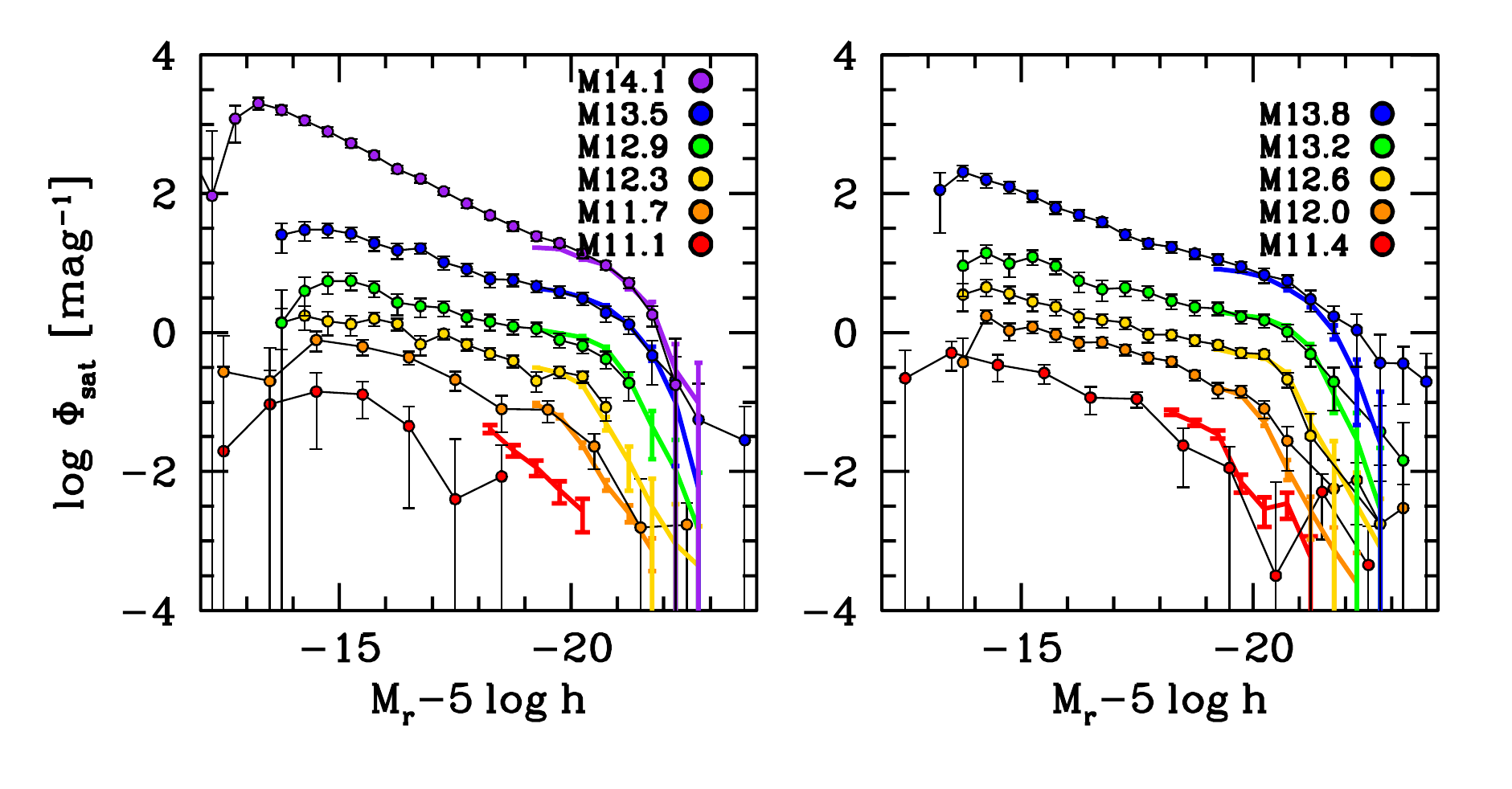}
\vspace{-.8cm}
\caption{ \label{lumfunc} Conditional luminosity functions for
  satellites within spectroscopic groups. Results are split into two
  panels to avoid crowding. The filled circles show results using the
  DLIS imaging data with background subtraction. The thick colored lines
  show the results using the spectroscopic redshifts obtained by
  matching the DLIS galaxies with SDSS spectra. Errors are from
  bootstrap resampling on the sample of groups. The key shows the
  value of $\log\mhalo$. Error bars are obtained by bootstrap
  resampling on the set of groups. }
\end{figure*}

\begin{figure*}
\includegraphics[width=7in ]{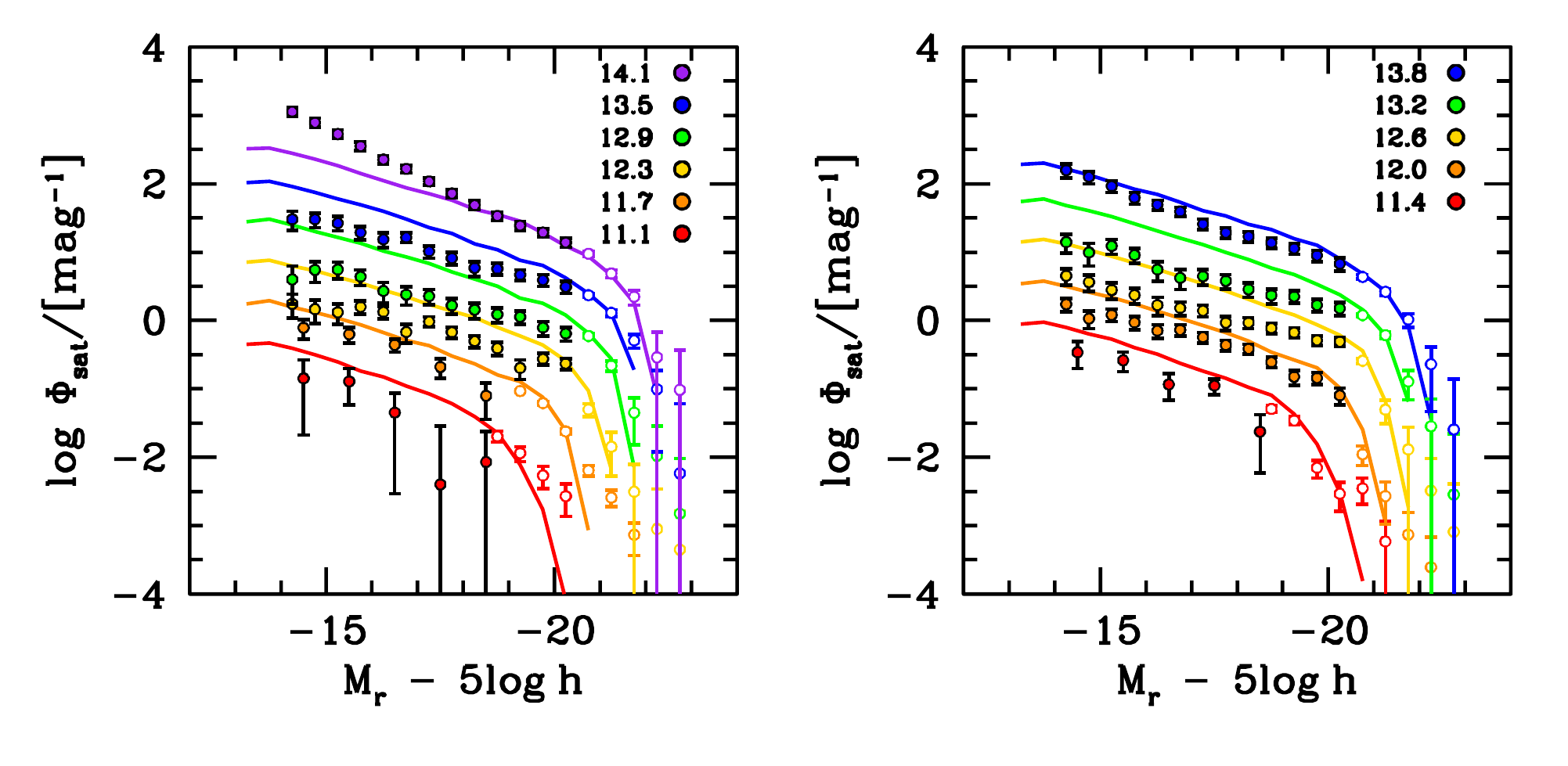}
\vspace{-.8cm}
\caption{ \label{clf_sham} Same as Figure \ref{lumfunc}, but now
  comparing the CLF measurements to abundance matching predictions of
  the C125 simulation. Solid curves show the N-body
  predictions. Circles indicate the DLIS measurements, with open circles
  being measurements using the DLIS galaxies with SDSS spectra,
  restricted to the magnitude range where the sample is complete.  }
\end{figure*}

\subsection{Conditional Luminosity Functions}

Figure \ref{lumfunc} presents our measurements of the CLFs within the
first and third volume-limited samples listed in Table 1. To avoid
crowding, we split the CLFs into two panels. In each panel, the
connected circles show the results from our method of stacking DLIS data
and subtracting off the background contribution estimated in annuli
around each halo. Errors are estimated by bootstrap resampling on the
sample of central galaxies. The thick colored curves show results from
DLIS galaxies that are brighter than the SDSS flux limit, and thus have
spectroscopic redshifts. For each DLIS galaxy with a redshift, we
calculate the probability that is is a satellite within a group in our
SDSS group catalogs using the same procedure that the group catalogs
were constructed in the first place (see Appendix A in
\citealt{tinker_etal:11}). These two independent methods of estimating
$\phisat$ show excellent agreement in the regions they overlap. The
spectroscopic sample is more efficient at removing extremely bright
galaxies that are spuriously counted as satellites, but the overall
amplitudes and Schechter-function break-points agree between the two
measurements. 

Figure \ref{clf_sham} compares the CLF measurements to abundance
matching predictions. Here, the data points are a composite of the
imaging results and the spectroscopic results. Overall, the agreement
is reasonable. The overall scaling of $\phisat$ with halo mass is in
good agreement (we will show integrated values of $\lsat$ in the next
section), and the abundance matching predictions match the observed
CLFs at the high-luminosity end. The primary disagreement between the
N-body prediction and the measurements is in the power-law slope at
faint luminosities, which is shallower in the data than in the
abundance matching model. This could be due to the possible artificial
disruption of satellites discussed with Figure \ref{radial_profile},
which would preferentially impact lower-mass subhalos, or it could be
that the extrapolation of our particular implementation of the
abundance matching model to such low luminosities is no longer
valid. We will address these questions in more detail in a subsequent
paper.

Previous measurements of the faint CLF in the Milky Way and M31 by
\cite{strigari_wechsler:12} found a cumulative number of satellites of
2.4 and 3.1, respectively, down to our magnitude limit of $-14$. At
$10^{12}$ $\msol$, our measurements yield a cumulative number of 2.9
satellites, which is in excellent agreement with these independent
results.

In Appendix \ref{s.app_clf} we present fitting functions for the
measured CLFs, where the parameters of a double Schechter function are
presented as functions of halo mass.

\begin{figure*}
\includegraphics[width=6in ]{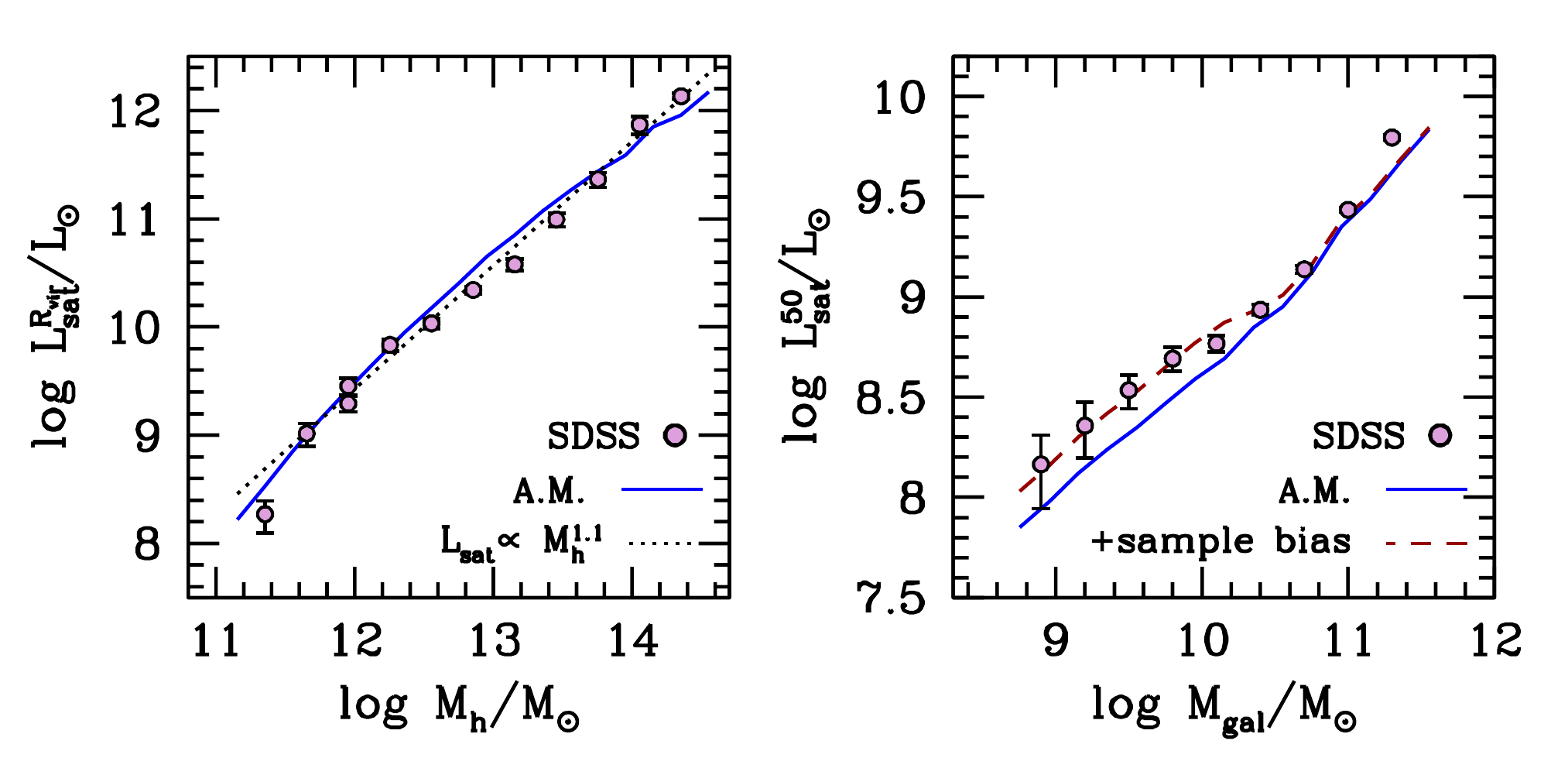}
\vspace{-0.3cm}
\caption{ \label{lsat_compare} {\it Left-hand Panel:} A comparison
  between $\lsatr$ measured around SDSS central galaxies and $\lsatr$
  predicted in $\lcdm$ simulations combined with abundance matching
  (``A.M''). The dotted line is the best-fit power law to the SDSS
  measurements, indicating that $\lsatr$ scales roughly linearly with
  $\mhalo$. {\it Right-hand Panel:} Same as the opposite panel, but
  only now binning by central galaxy stellar mass. The dashed curve
  shows the prediction of abundance matching after adding the expected
  bias in $\lsatx$ yielded by the central galaxy finder, calibrated in
  Paper I.}
\end{figure*}

\subsection{Total Satellite Luminosity}

Figure \ref{lsat_compare} compares the total integrated satellite
luminosity, $\lsat$, measured in the data to our abundance matching
predictions. All results use our fiducial limiting magnitude of
$M_r-5\log h = -14$. The left-hand panel shows the comparison for
$\lsatr$. The SDSS sample uses the volume-limited group catalogs, with
$\psat<0.1$ and halo masses estimated by the group finder. Error bars
are from bootstrap resampling on the sample of groups. The dotted
black line is a power-law fit to these data, which yields a scaling of
$\lsatr\propto \mhalo^{1.1}$. By a simple $\chi^2$ statistic, a
power-law is a statistically acceptable description of these data. The
solid blue curve shows the abundance matching prediction. The $\lcdm$
prediction scales close to a power law, being steeper at
$\mhalo<10^{13}$ $\msol$ and shallower above this scale.

The right-hand panel of Figure \ref{lsat_compare} compares our
measurements of $\lsatx$ to the abundance matching predictions. The
observational results use the full flux-limited SDSS catalog, with
central galaxies identified using the central-finding
algorithm. Results are shown as a function of $\mgal$. The blue curve
once again shows our abundance matching prediction. At low stellar
masses, $\mgal<10^{10.5}$ $\msol$, the measurements are above the
predictions by about 0.2 dex. As demonstrated in Paper I, the
central-finding algorithm induces some impurities in the sample of
central galaxies. Using mocks, the impact of these impurities on
$\lsatx$ was quantified. The dashed curve in this panel adds the bias
on $\lsatx$ to the abundance matching prediction, brining the theory
and observations into near perfect agreement.

\subsection{Comparison to Weak Lensing Measurements}

\begin{figure*}
\includegraphics[width=3.3in ]{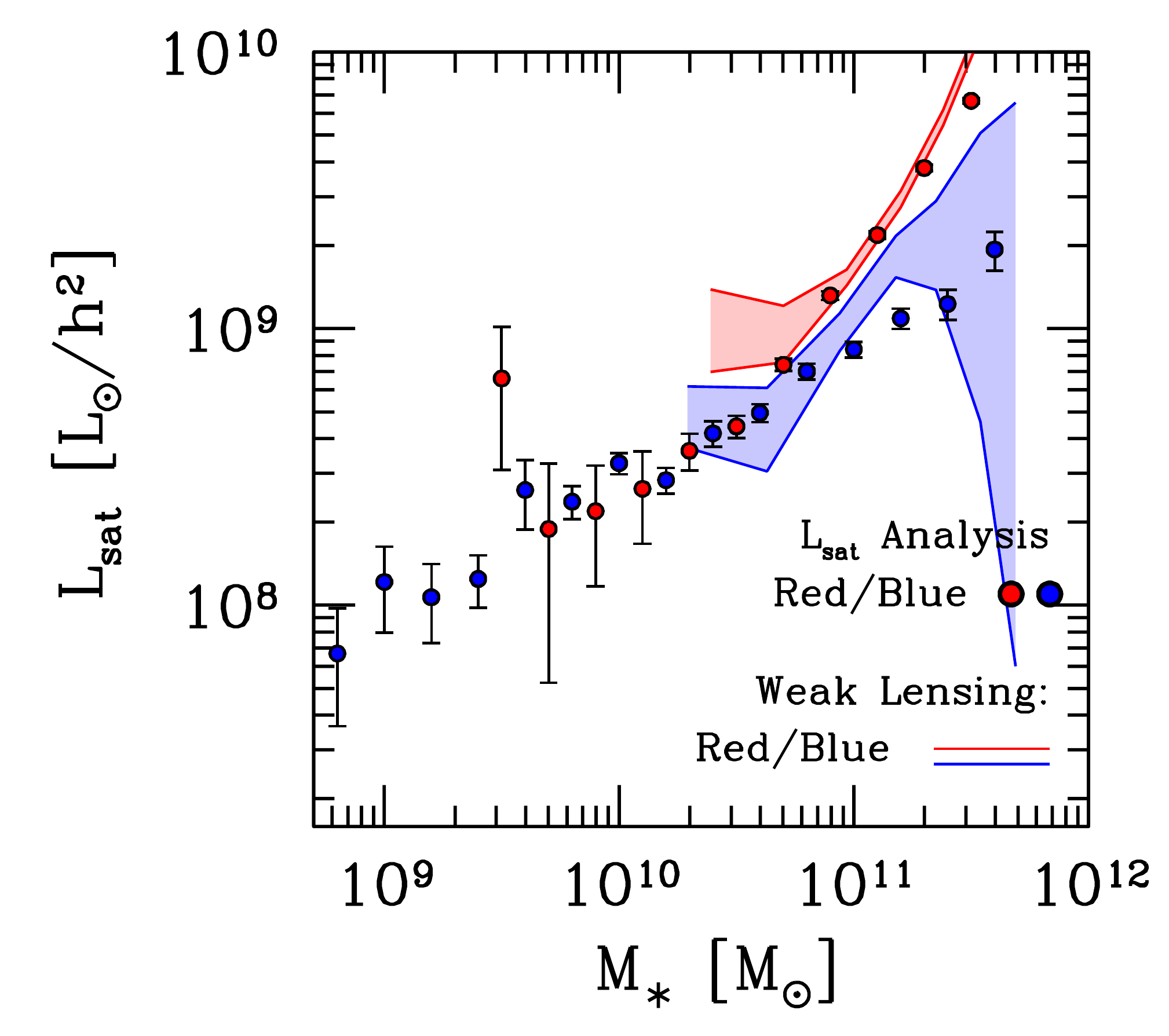}
\includegraphics[width=3.2in ]{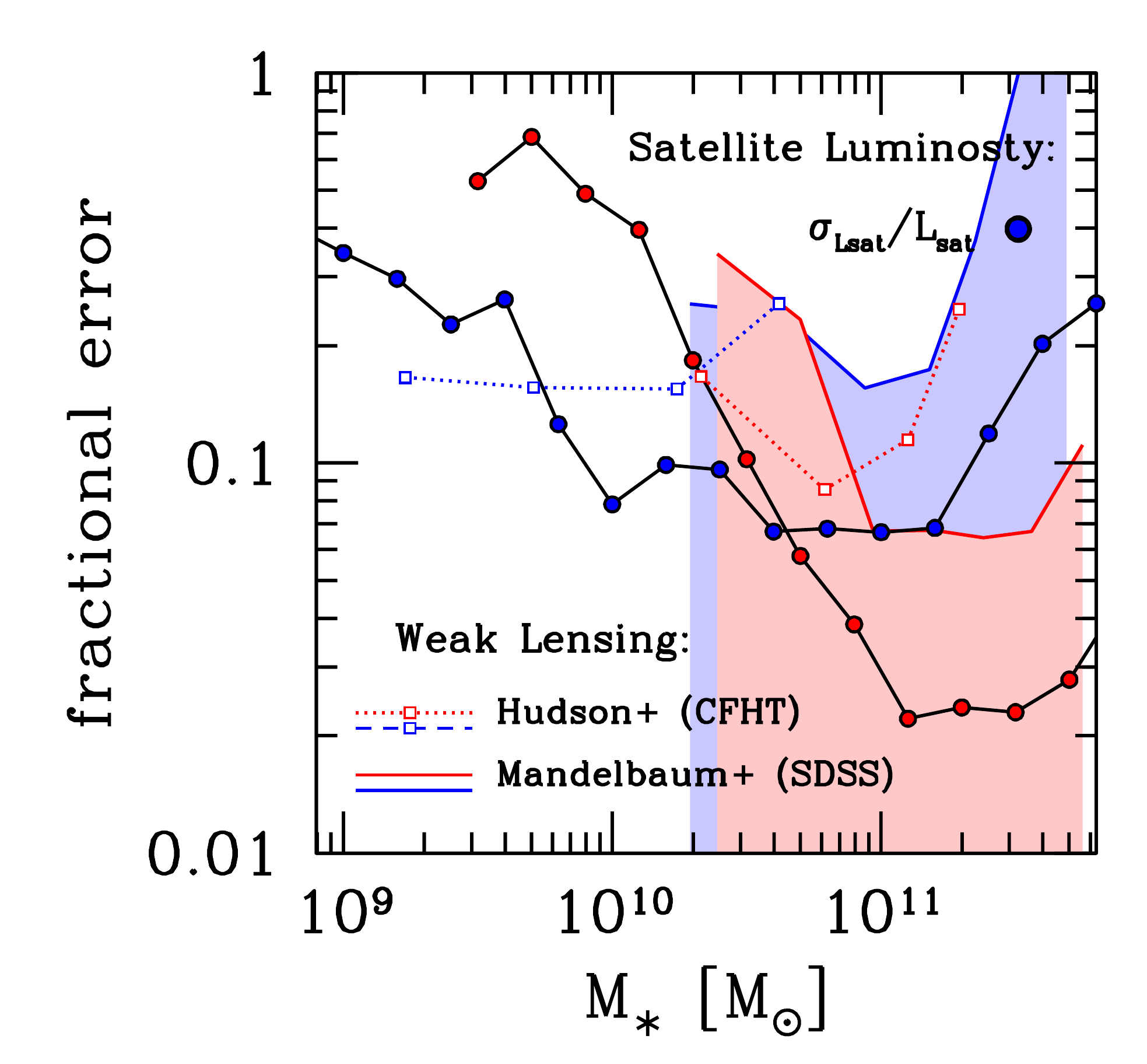}
\caption{ \label{errors} {\it Left Panel:} Comparing our $\lsatx$
  measurements to weak lensing measurements of SDSS central galaxies
  from \citet{mandelbaum_etal:16}. Points with errors show our $\lsatx$
  measurements. At each $\mgal$, galaxies are divided by the same
  color cut as used in Mandelbaum, $g-r>0.8$. The shaded regions
  indicate the weak lensing constraints. We convert the halo mass
  estimates of Mandelbaum to $\lsatx$ values using the abundance
  matching results in Figure \ref{lsat_theory}. {\it Left Panel:} A
  comparison of the errors between $\lsatx$ and weak lensing. The
  filled circles are from our $\lsatx$ measurements, now broken into
  star-forming and quiescent (``Red/Blue'') samples. The shaded
  regions indicate the errors on weak lensing halo masses by
  \citet{mandelbaum_etal:16}, which is performed on the same SDSS
  sample as our $\lsatx$ results. The open squares connected by dotted
  lines show results from the CFHT legacy survey by
  \citet{hudson_etal:15}, a smaller area but deeper imaging. }
\end{figure*}

Figure \ref{errors} compares our $\lsatx$ measurements to weak lensing
halo mass estimates from \cite{mandelbaum_etal:16}. The Mandelbaum
results split the galaxy sample by color, with red and blue galaxies
separated at $g-r=0.8$. To facilitate a proper comparison, we convert
$\mhalo$ to $\lsatx$ using the abundance matching results in Figure
\ref{lsat_theory}. The weak lensing results are shown with the red and
blue shaded regions in the left-hand panel. Our $\lsatx$ measurements
are shown with the points with errors. Red and blue circles indicate
results for the red and blue subsamples, split with the same $g-r$
color cut. As with the weak lensing results, the $\lsat$ measurements
indicate that red and blue central galaxies live in halos of different
mass, and this difference gets larger as $\mgal$ increases. The
relative values of $\lsatx$ for the blue and red subsamples is in good
agreement between the two independent approaches.

The left side of Figure \ref{errors} also highlights how $\lsat$ is
complementary to weak lensing, in that the $\lsat$ measurements can be
mae to much lower values of $\mgal$, below the limiting mass for
lensing. At these lower masses, the $\lsatx$ values for the blue and
red subsamples converge, indicating that low-mass central galaxies on
the red sequence live in the same halos as star-forming central
galaxies of the same $\mgal$. 


The right-hand side of Figure \ref{errors} compares the errors on
$\lsatx$ to those obtained from weak lensing mass estimates. At low
$\mgal$, the error on red galaxies for both lensing and $\lsatx$ is
significantly higher due to the lower overall number of red
galaxies. This reverses at the same location for both methods,
$\mgal\sim 4\times 10^{10}$ $\msol$. But the errors on $\lsatx$ are
roughly a factor of five lower than those on the weak lensing
masses.  We also compare to results from the CFTH legacy survey from
\cite{hudson_etal:15}, representing a much smaller-area survey but
higher quality imaging. The result of this survey design leads to
better errors than the SDSS results at low $\mgal$, but the smaller
volume limits the upper mass limit at which robust measurements can be
obtained. But the errors on $\lsatx$ are still significantly lower
than those from CFHT at most masses.

\section{Summary}

In this paper we have presented $\lsat$ as a proxy for dark matter
halos. The approach is to measure the total luminosity in satellite
galaxies around a central galaxy. $\lsat$ is significantly more
sensitive to $\mhalo$ than the luminosity or mass of the central
galaxy, with a nearly linear dependence between $\lsatr$ and
$\mhalo$. To minimize any priors on the halo mass one would infer from
this method, we also explore $\lsat$ measured within fixed apertures
of 100 $\hkpc$ and 50 $\hkpc$. At $\mgal<10^{10.5}$ $\msol$ and
$\mhalo<10^{12.5}$ $\msol$, there is minimal impact on the scaling of
$\lsat$ when enforcing a fixed aperture. At larger masses, enforcing a
fixed aperture makes $\lsat$ less sensitive to halo mass than
estimating $\lsat$ using all satellites within the halo, but there is
a still a monotonic relationship between $\lsat$ and $\mgal$ at all
stellar masses.

A number of tests and comparisons demonstrate the robustness of our
approach:

\begin{itemize}
\item We find good agreement between theoretical predictions of
  $\lsat$ from abundance matching models, and our measurements of
  $\lsat$ around SDSS galaxies.
\item We find good agreement between the conditional luminosity
  functions based on our $\lsat$ approach using photometric data, and
  the CLF obtained from spectroscopic-only data, where the two
  overlap.
\item We find good agreement between $\lsat$ and weak lensing results
  for the relative halo masses of blue and red SDSS galaxies at fixed
  $\mgal$. 
\item Given the additional signal-to-noise of the $\lsat$ method, we
  are able to extend this comparison of red and blue galaxies to
  galaxies that are more than an order of magnitude lower mass.
\end{itemize}
  
\noindent Additionally, at higher stellar masses, the precision of the
$\lsat$ method allows us to compare galaxies not just in divisions of
red and blue, but then to further fine-bin these divisions based on
other galaxy properties. \cite{alpaslan_tinker:19}, a companion paper
to this manuscript, presents these initial results from SDSS.

The main theoretical uncertainty in using $\lsat$ as a halo mass proxy
is the degeneracy of $\lsat$ on halo formation history; late-forming
halos have more substructure and thus larger amounts of satellite
luminosity. However, this correlation makes a distinct prediction for
the clustering and large-scale environments of central galaxies. For
example, \cite{tinker_etal:18_p3}, looking at galaxies on the
star-forming main sequence, showed that star formation rate correlated
with large-scale environment at fixed stellar mass: above-average
star-forming galaxies live in below-average large-scale
densities. These results were consistent with a model in which galaxy
star-formation rate was correlated with dark matter halo accretion
rate. Thus, if $\lsat$ correlates with star formation rate, the
measurements of \cite{tinker_etal:18_p3} can break the degeneracy, and
quantify how much of the $\lsat$ variation is due to a change in
$\mhalo$ how much is due to a correlation with $\zhalf$. But this is
only for star-forming galaxies. In contrast, when dividing galaxies
into star-forming and quiescent samples, the quenched fraction of
central galaxies is {\it independent} of environment
(\citealt{tinker_etal:08_voids, tinker_etal:17_p1, tinker_etal:18_p2,
  peng_etal:10, zu_mandelbaum:16, zu_mandelbaum:18, elucid4}). Thus,
if there are differences in $\lsat$ between red and blue galaxies,
this is truly due to differences in $\mhalo$, and not correlated with
$\zhalf$. This approach can be extended to any galaxy property one
wishes to probe, including galaxy size, velocity dispersion, and
morphology.

The primary observational systematic is in the definition of the
sample of central galaxies. Impurities in this sample always go in the
direction of increasing $\lsat$ at fixed $\mgal$. No method of
identifying central galaxies will be perfect, but the two methods used
here have minimal impact on $\lsat$, increasing it by 0.1 to 0.2
dex. The primary concern is when measuring relative values of $\lsat$
when splitting up galaxies by secondary properties at fixed
$\mgal$. If the secondary property correlates with satellite fraction,
the impact of impurities will have a differential effect on the
relative values of $\lsat$. The division that should maximize this
error---splitting the sample based on color or star-formation
bimodality---still only produced a bias of 0.1-0.2 dex. The results
from our mock galaxy tests can also be used to correct observational
results, or set systematic error bars. 

The $\lsat$ technique opens a new window into the galaxy-halo
connection. Many of the outstanding issues in this relationship can be
addressed with $\lsat$.
In a recent review of the galaxy-halo connection,
\cite{wechsler_tinker:18} highlight the need for additional data to
constrain the relationship between halo mass and secondary galaxy
properties, including galaxy color and galaxy size. However, these
secondary relationships are difficult to detect through previous
methods.  Given these open questions, the $\lsat$ technique has many
applications that are complementary with direct probes of halo mass.

\section*{Acknowledgement}
The simulations used in this study were produced with computational resources at SLAC National Accelerator Laboratory, a U.S.\ Department of Energy Office; YYM and RHW thank Matthew R. Becker for creating these simulations, and thank the support of the SLAC computational team. 

YYM was supported by the Samuel P.\ Langley PITT PACC Postdoctoral Fellowship, and by NASA through the NASA Hubble Fellowship grant no.\ HST-HF2-51441.001 awarded by the Space Telescope Science Institute, which is operated by the Association of Universities for Research in Astronomy, Incorporated, under NASA contract NAS5-26555.

This research used resources of the National Energy Research Scientific Computing Center (NERSC), a U.S. Department of Energy Office of Science User Facility operated under Contract No. DE-AC02- 05CH11231.

The Legacy Surveys consist of three individual and complementary projects: the Dark Energy Camera Legacy Survey (DECaLS; NOAO Proposal ID \# 2014B-0404; PIs: David Schlegel and Arjun Dey), the Beijing-Arizona Sky Survey (BASS; NOAO Proposal ID \# 2015A-0801; PIs: Zhou Xu and Xiaohui Fan), and the Mayall z-band Legacy Survey (MzLS; NOAO Proposal ID \# 2016A-0453; PI: Arjun Dey). DECaLS, BASS and MzLS together include data obtained, respectively, at the Blanco telescope, Cerro Tololo Inter-American Observatory, National Optical Astronomy Observatory (NOAO); the Bok telescope, Steward Observatory, University of Arizona; and the Mayall telescope, Kitt Peak National Observatory, NOAO. The Legacy Surveys project is honored to be permitted to conduct astronomical research on Iolkam Du'ag (Kitt Peak), a mountain with particular significance to the Tohono O'odham Nation.

NOAO is operated by the Association of Universities for Research in Astronomy (AURA) under a cooperative agreement with the National Science Foundation.

This project used data obtained with the Dark Energy Camera (DECam), which was constructed by the Dark Energy Survey (DES) collaboration. Funding for the DES Projects has been provided by the U.S. Department of Energy, the U.S. National Science Foundation, the Ministry of Science and Education of Spain, the Science and Technology Facilities Council of the United Kingdom, the Higher Education Funding Council for England, the National Center for Supercomputing Applications at the University of Illinois at Urbana-Champaign, the Kavli Institute of Cosmological Physics at the University of Chicago, Center for Cosmology and Astro-Particle Physics at the Ohio State University, the Mitchell Institute for Fundamental Physics and Astronomy at Texas A\&M University, Financiadora de Estudos e Projetos, Fundacao Carlos Chagas Filho de Amparo, Financiadora de Estudos e Projetos, Fundacao Carlos Chagas Filho de Amparo a Pesquisa do Estado do Rio de Janeiro, Conselho Nacional de Desenvolvimento Cientifico e Tecnologico and the Ministerio da Ciencia, Tecnologia e Inovacao, the Deutsche Forschungsgemeinschaft and the Collaborating Institutions in the Dark Energy Survey. The Collaborating Institutions are Argonne National Laboratory, the University of California at Santa Cruz, the University of Cambridge, Centro de Investigaciones Energeticas, Medioambientales y Tecnologicas-Madrid, the University of Chicago, University College London, the DES-Brazil Consortium, the University of Edinburgh, the Eidgenossische Technische Hochschule (ETH) Zurich, Fermi National Accelerator Laboratory, the University of Illinois at Urbana-Champaign, the Institut de Ciencies de l'Espai (IEEC/CSIC), the Institut de Fisica d'Altes Energies, Lawrence Berkeley National Laboratory, the Ludwig-Maximilians Universitat Munchen and the associated Excellence Cluster Universe, the University of Michigan, the National Optical Astronomy Observatory, the University of Nottingham, the Ohio State University, the University of Pennsylvania, the University of Portsmouth, SLAC National Accelerator Laboratory, Stanford University, the University of Sussex, and Texas A\&M University.

BASS is a key project of the Telescope Access Program (TAP), which has been funded by the National Astronomical Observatories of China, the Chinese Academy of Sciences (the Strategic Priority Research Program "The Emergence of Cosmological Structures" Grant \# XDB09000000), and the Special Fund for Astronomy from the Ministry of Finance. The BASS is also supported by the External Cooperation Program of Chinese Academy of Sciences (Grant \# 114A11KYSB20160057), and Chinese National Natural Science Foundation (Grant \# 11433005).

The Legacy Survey team makes use of data products from the Near-Earth Object Wide-field Infrared Survey Explorer (NEOWISE), which is a project of the Jet Propulsion Laboratory/California Institute of Technology. NEOWISE is funded by the National Aeronautics and Space Administration.

The Legacy Surveys imaging of the DESI footprint is supported by the Director, Office of Science, Office of High Energy Physics of the U.S. Department of Energy under Contract No. DE-AC02-05CH1123, by the National Energy Research Scientific Computing Center, a DOE Office of Science User Facility under the same contract; and by the U.S. National Science Foundation, Division of Astronomical Sciences under Contract No. AST-0950945 to NOAO.

\appendix
\section{Finding Central Galaxies}
\label{s.app_cenfinder}

\begin{figure*}
\includegraphics[width=6in]{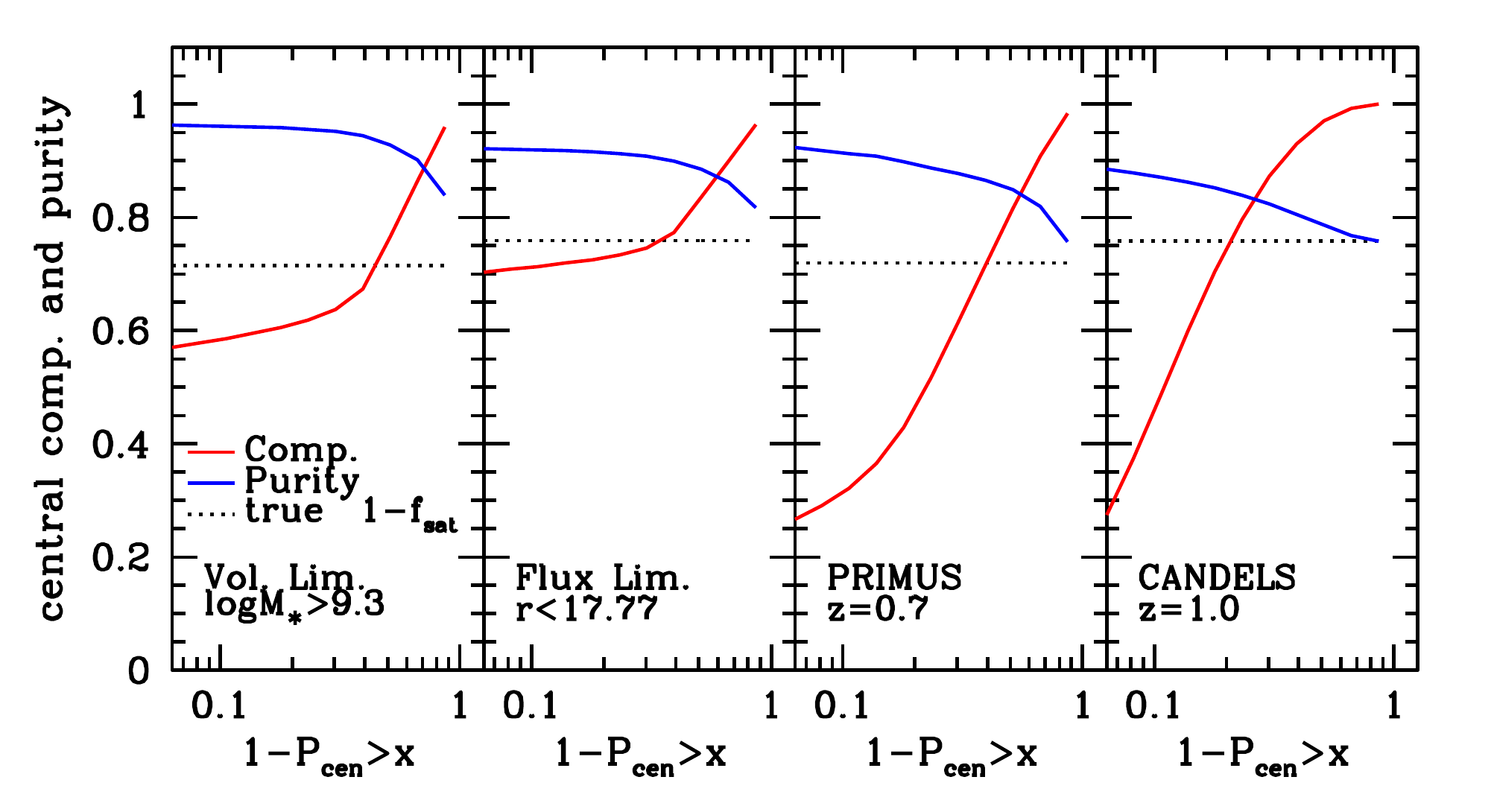}
\caption{ \label{cenfinder} The purity and completeness of central
  galaxies identified by our central-finding algorithm. The solid
  curves show the results as a function of $\psat$ threshold. The
  dotted line shows the true fraction of central galaxies in each mock
  galaxy catalog. From left to right, the panels are: a volume-limited
  sample of galaxies, complete to a stellar mass of $\mgal=10^{9.3}$
  $\msol$. A flux-limited sample of galaxies with the same flux limit
  as the SDSS main galaxy sample, and the same $n(z)$. A stellar-mass
  complete sample of galaxies at $z=0.7$ with redshift errors
  comparable to those in the PRIMUS survey. A stellar-mass complete
  sample of galaxies at $z=1.0$ with photometric redshift errors set
  to equal those from the CANDELS survey. }
\end{figure*}

Here we present the details of our central-finding algorithm. The
purpose of this algorithm is to determine if a galaxy is likely to be
a central, but without attempting to determine the mass of its host
halo. The basic approach is very similar to running to the
initialization stage of our group finder, which works on inverse
abundance matching---i.e., rather than putting mock galaxies within
N-body dark matter halos, we put halos around observed galaxies by the
same abundance matching calculation:

\begin{equation}
\label{e.sham}
\int_{\mgal}^\infty \Phi(\mgal^\prime)\,d\mgal^\prime = 
\int_{\mhalo}^\infty n(\mhalo^\prime)\,d \mhalo^\prime, 
\end{equation}

\noindent where $\Phi(\mgal)$ is the observed stellar mass function
and $n(\mhalo)$ is the mass function of dark matter halos. Equation
(\ref{e.sham}) is the simplest form of abundance matching, assuming no
scatter between $\mgal$ and $\mhalo$. The first step in our group
finder is to use equation (\ref{e.sham}) as a first guess of the halos
around each galaxy, regardless of whether they are centrals or
satellites, then begin the process of determining the probability that
each galaxy is a satellite within a larger dark matter halos. In the
group finder, the halo mass is itself abundance matched onto the {\it
  total group stellar mass}, and the entire sample is iterated to
convergence. 

In our central finder, we make the approach even more simple and
flexible by using pre-tabulated abundance matching relations between
$\mhalo$ and $\mgal$. In our fiducial approach, we use the relations
tabulated in \cite{behroozi_etal:13}, which quantify the
stellar-to-halo mass relation from $z=0$ to $z=8$. The use of
pre-tabulated relations means that it is no longer necessary to
perform the abundance matching on volume-limited samples of galaxies;
a $10^{11}$ $\msol$ central galaxy at $z=0.02$, which likely has 10's
of satellites within the flux-limited SDSS main galaxy sample, is
assigned the same halo mass at the same galaxy observed at $z=0.15$,
where it likely has no satellite galaxies. 

Just as in the group finder, the probability that a galaxy is a
satellite in a larger dark matter halo is given by:

\begin{equation}
\pcen = \left(1+P_{R_p}P_{\Delta z}/B\right)^{-1}
\end{equation}

\noindent where $P_{R_p}$ is the probability at a given projected
separation from the center of the halo, and $P_{\Delta z}$ is the
probability at a given line-of-sight separation from the redshift of
the halo. $B$ is a constant determined from calibration on mock galaxy
samples, set to be $B=10$. The former is given by the projected NFW
density profile (\citealt{nfw:97}), while the latter assumes a
Gaussian probability distribution function with width given by the
virial velocity dispersion of the host halo. Further details can be
found in Appendix A of \cite{tinker_etal:11} or \cite{yang_etal:05},
on which our algorithm is based.

The drawback to using the inverse abundance-matching approach is that
the halos assigned to galaxies will be biased, given the fact that
asymmetric scatter yields different mean relationships of $\langle
\mgal | \mhalo \rangle$ and $\langle
\mhalo | \mgal \rangle$. However, this the error this generally
accrues is to overestimate the masses of host halos, thus making it a
conservative approach to determining if nearby galaxies are
satellites. The other drawback to this approach is that stellar mass
estimates can differ widely, thus the estimate used in
\cite{behroozi_etal:13} may differ from that used in a given sample of
galaxies. However, in our tests we find that this yields minimal bias.

Figure \ref{cenfinder} shows the results of our central finder when
applied to four different galaxy mocks. As a baseline, the left-hand
panel shows results when applied to a volume-limited sample of $z=0$
galaxies, complete down to a stellar mass of $\mgal=10^{9.3}$
$\msol$. The stellar mass function used to put stellar mass into the
halos is the PCA stellar mass function. Normally, one would use the
full group finder on a volume-limited sample, but this test provides a
good comparison for how results degrade in less optimal survey
samples. The second panel is a flux-limited sample with same redshift
distribution as the SDSS main galaxy sample. For $\psat<0.3$, the
sample of central galaxies identified by the algorithm has a purity of
$>90\%$ and a completeness of around 70\%. 

The right two panels show the algorithm as applied to mock samples
that have significantly degraded redshift information. Here, we
replace the host halo velocity dispersion used the halo in $P_{\Delta
  z}$ with the average error on the redshift. To construct mock
samples we use the Buzzard mocks of \cite{buzzard}.  The rightmost
panel is a mock sample comparable to the CANDELS survey at $z=1.0$
(\citealt{candels}); it is complete down to a stellar mass of
$\mgal=10^{9.5}$ with a photometric redshift error of
$\sigma_z/(1+z)=0.033$. The next rightmost panel is an approximation
of the redshift accuracy and redshift range of the PRIMUS survey
(\citealt{primus}), with $z\approx 0.7$ and
$\sigma_z/(1+z)=0.005$. For both of these mock surveys, the algorithm
yields central galaxy samples that are complete to $\sim 90\%$ at
$\psat<0.1$. However, this comes at a significant cost of the
completeness of the sample, which dips below 50\% at this $\psat$
threshold. 

This code and a short instruction manual is made publicly
available\footnote{{\tt https://github.com/jltinker/IsolationCriterion}}.

\section{Fitting the Conditional Luminosity Functions}
\label{s.app_clf}

\begin{figure*}
\includegraphics[width=7in ]{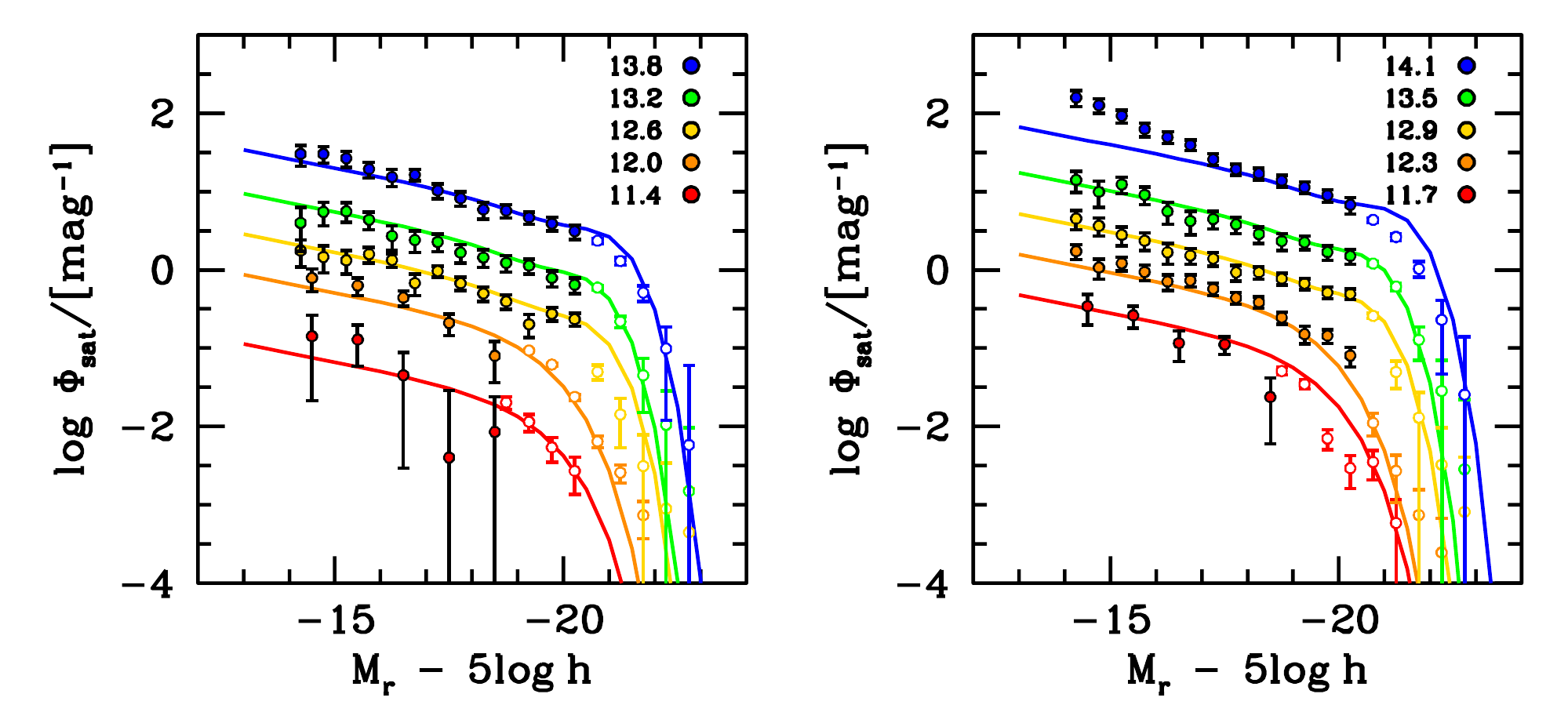}
\vspace{-.5cm}
\caption{ \label{clf_fits} Same as Figure \ref{lumfunc}, but now
  comparing the CLF measurements to double-Schechter function
  fits. The parameters of the fitting function vary with
  $\log\mhalo$. The fits are optimized for halo mass bins
  $\log\mhalo<14.1$. Larger halos suffer from insufficient background
  subtraction at the very faint end of the CLF. }
\end{figure*}

We model the conditional luminosity function of satellites within
halos using the modified Schechter function employed by
\cite{blanton_etal:05} to describe the luminosity function of
low-luminosity galaxies in the SDSS. This fitting function takes the
form

\begin{equation}
\begin{split}
\Phi(M) =  0.4 \ln 10 dM
  \exp\left(-10^{-0.4(M-M_{\ast})}\right) \\\left[
\phi_{\ast,1} 
10^{-0.4 \left( M-M_{\ast} \right)(\alpha_1+1)} +
\phi_{\ast,2} 
10^{-0.4 \left( M-M_{\ast} \right)(\alpha_2+1)}
  \right].
\end{split}
\end{equation}

\noindent The motivation for this function in \cite{blanton_etal:05}
is to better model an upturn in the luminosity function at magnitudes
fainter than $M_r=-18$. A similar upturn is seen in the CLFs in higher
mass halos in our results in Figure \ref{lumfunc}. We fix the values
of the power-law indices to be of $\alpha_1=1$ and $\alpha_2=-1.28$,
with the latter parameter fixing the faint-end slope to be the same
for all halos. We construct fitting functions for how the other
parameters of the modified Schechter function depend on halo mass;






\begin{equation}
\label{e.phi1}
\phi_{\ast,1} = \left\{ \begin{array}{ll}
    0.98x - 12.85 & {\rm if\ \ } x>12.5 \\
		0 & {\rm if\ \ } x\le 12.5  \,\, ,\\
                \end{array}
\right.
\end{equation}

\noindent where $x\equiv \log M_h$, 

\begin{equation}
\phi_{\ast,2} = \left\{ \begin{array}{ll}
    0.86x - 11.10  & {\rm if\ \ } x>11.7 \\
		2.13x - 26 & {\rm if\ \ } x\le 11.7  \,\, ,\\
                \end{array}
\right.
\end{equation}

\noindent and

\begin{equation}
M_{\ast} = \left\{ \begin{array}{ll}
    -0.99x - 6.36  & {\rm if\ \ } x>13.3 \\
		-19.58 & {\rm if\ \ } x\le 13.3  \,\, .\\
                \end{array}
\right.
\end{equation}

Equation (\ref{e.phi1}) implies that, for halos less massive than
$10^{12.5}$ $\msol$, a single power-law Schechter function is
sufficient for modeling the CLF. The results of the fitting functions
are shown in Figure \ref{clf_fits}. The fitting parameters themselves
were obtained for the results at $\mhalo\le 10^{13.8}$ $\msol$. At
higher halo masses, the faint end of the CLF deviates significantly
from the trends seen in lower mass halos. This is likely due to
insufficient background subtraction for the most massive halos, and
this effect is also seen in our tests with mock galaxy samples (see
Figure 8 in Paper I).

\bibliography{../risa}

\label{lastpage}

\end{document}